\documentclass[12pt]{spieman}  
\usepackage{amsmath,amsfonts,amssymb}
\usepackage{graphicx}
\usepackage{setspace}
\usepackage{tocloft}
\usepackage{xspace}

\newcommand{\ie}{{\it i.e.}\xspace}
\newcommand{\eg}{{\it e.g.}\xspace}

\newcommand{\etc}{{\it etc.}\xspace}
\newcommand{\etal}{{\it et al.}\xspace}

\newcommand{\HgCdTeL}{Hg$_{1-x}$Cd$_x$Te\xspace}
\newcommand{\oxygen}{$\rm O_2$\xspace}
\newcommand{\ozone}{$\rm O_3$\xspace}
\newcommand{\methane}{$\rm CH_4$\xspace}
\newcommand{\water}{$\rm H_2 O$\xspace}
\newcommand{\cotwo}{$\rm CO_2$\xspace}
\newcommand{\carbonmonoxide}{$\rm CO$\xspace}

\title{Detectors and cooling technology for direct spectroscopic biosignature characterization\footnote{Please cite as Bernard J. Rauscher, Edgar R. Canavan, Samuel H. Moseley, John E. Sadleir, Thomas Stevenson, "Detectors and cooling technology for direct spectroscopic biosignature characterization," J. Astron. Telesc. Instrum. Syst. 2(4), 041212 (2016), doi: 10.1117/1.JATIS.2.4.041212. }}

\author[a,*]{Bernard J. Rauscher}
\author[b]{Edgar R. Canavan}
\author[a]{S.H. Moseley}
\author[c]{John E. Sadleir}
\author[c]{\\Thomas Stevenson}
\affil[a]{NASA Goddard Space Flight Center, Observational Cosmology Laboratory, Greenbelt, USA, 20771}
\affil[b]{NASA Goddard Space Flight Center, Cryogenics and Fluids Branch, Greenbelt, USA, 20771}
\affil[c]{NASA Goddard Space Flight Center, Detector Systems Branch, Greenbelt, USA, 20771}

\cftpagenumbersoff{figure}
\cftpagenumbersoff{table} 
\begin{document} 
\maketitle

\begin{abstract}
Direct spectroscopic biosignature characterization (hereafter ``biosignature characterization'') will be a major focus for future space observatories equipped with coronagraphs or starshades. Our aim in this article is to provide an introduction to potential detector and cooling technologies for biosignature characterization. We begin by reviewing the needs. These include nearly noiseless photon detection at flux levels as low as $<0.001~\textrm{photons}~s^{-1}~\textrm{pixel}^{-1}$ in the visible and near-IR. We then discuss potential areas for further testing and/or development to meet these needs using non-cryogenic detectors (EMCCD, HgCdTe array, HgCdTe APD array), and cryogenic single photon detectors (MKID arrays and TES microcalorimeter arrays). Non-cryogenic detectors are compatible with the passive cooling that is strongly preferred by coronagraphic missions, but would add non-negligible noise. Cryogenic detectors would require active cooling, but in return deliver nearly quantum limited performance. Based on the flight dynamics of past NASA missions, we discuss reasonable vibration expectations for a large UV-Optical-IR space telescope (LUVOIR) and preliminary cooling concepts that could potentially fit into a vibration budget without being the largest element. We believe that a cooler that meets the stringent vibration needs of a LUVOIR is also likely to meet those of a starshade-based Habitable Exoplanet Imaging Mission. 
\end{abstract}


{\noindent \footnotesize\textbf{*}Bernard J. Rauscher,  \linkable{Bernard.J.Rauscher@nasa.gov} }

\begin{spacing}{1}   

\section{Introduction}\label{sec:intro}

The search for life on other worlds looms large in NASA's 30-year strategic vision.\cite{2014arXiv1401.3741K} Already, several mission concept studies are either completed or underway that would use a larger than 8 meter aperture UV-Optical-IR space telescope equipped with a coronagraph or starshade to characterize potentially habitable exoEarths (\eg ATLAST, HDST, LUVOIR).\footnote{The acronyms stand for: LUVOIR = Large UV-Optical-IR Surveyor,\cite{2014arXiv1401.3741K} ATLAST = Advanced Technology Large Aperture Space Telescope,\cite{Bolcar:2015jh,Rauscher:2015hba,Rioux:2015hn} and HDST = High Definition Space Telescope\cite{2015arXiv150704779D}.} Alternatively, smaller starshade-based Habitable-Exoplanet Imaging Mission concepts exist.\cite{NASATownhall:2015tm} All would benefit from better visible and near-IR (VISIR; $\lambda=400~\textrm{nm} - 2.5~\mu\textrm{m}$) detectors than exist today. Moreover, because of different overall system design considerations, different solutions may turn out to be optimal depending on whether a mission is coronagraph or starshade based. Our aim in this article is to discuss a short list of technologies that we believe to be potentially capable of biosignature characterization for either coronagraph or starshade missions.

Once a rocky exoplanet in the habitable zone has been found, biosignature characterization will be the primary tool for determining whether we think it harbors life. Biosignature characterization uses moderate resolution spectroscopy, $R=\lambda/\Delta\lambda>100$, to study atmospheric spectral features that are thought to be necessary for life, or that can be created by it (\eg \water, \oxygen, \ozone, \methane, \cotwo). We discuss these biosignatures in more detail in Sec.~\ref{sec:biosig} and the spectral resolution requirements for observing them in Sec.~\ref{sec:resolution}. Even using a very large space telescope, biosignature characterization is extremely photon starved. Ultra low noise detectors are needed, and true energy resolving single photon detectors would be preferred if they could be had without the vibration that is associated with conventional cryocoolers.

Our aim in this article is to provide an introduction to the detector needs for biosignature characterization, and some of the emerging technologies that we believe hold promise for meeting them within the next decade. The technologies fall into two broad categories: (1) low noise detectors (including ``photon counting'') that are compatible with passive cooling and (2) true energy resolving single photon detectors that require active cooling.

We draw a clear distinction between photon counting low noise detectors and single photon detectors. A photon counting detector is able to resolve individual photons, although the detection process still adds significant noise. For example, many kinds of photon counting detector have significant dark current and spurious charge generation at the ultra low flux levels that are encountered during biosignature characterization. A single photon detector, on the other hand, provides essentially noiseless detection of light. Noise in the single photon detectors discussed here manifests as an uncertainty in the energy of a detected photon rather than an uncertainty in the number of photons.

The low noise detectors include electron multiplying charge coupled devices (EMCCD) for the visible and HgCdTe photodiode and avalanche photodiode (APD) arrays for the near-IR. With targeted investment, we believe that all can be  improved beyond today's state of the art. Sec.~\ref{sec:btb} describes a low risk but evolutionary payoff route to improving these existing non-cryogenic detectors for use with conventional spectrographs. One advantage of this approach is that it completely retires the risks, cost, and complexities associated with a cryocooler. The disadvantages include increased noise and the need for dispersive spectrograph optics.

The single photon detectors that we discuss are based on thin superconducting films and operate at $\rm T\approx 100~mK$. Cryogenic cooling is required to achieve these temperatures. In return for cryogenic cooling, single photon detectors promise noiseless (in the conventional astronomy sense), nearly quantum limited photon detection with built in energy resolution. The built in energy resolution offers the tantalizing prospect of non-dispersive imaging spectrometry, thereby eliminating most spectrograph optics. In this article, we focus on two single photon detectors that have already been used for astronomy and that offer the potential for multiplexing up to sufficiently large formats. These are microwave kinetic inductance device (MKID) arrays and transition-edge sensor (TES) microcalorimeter arrays. Sec.~\ref{sec:spd} discusses a path forward using single photon detectors that offers the potential for nearly quantum limited detector performance and non-dispersive imaging spectrometry if the cooling challenges can be met.

Cryogenic cooling in the context of LUVOIR brings its own challenges. High performance space coronagraphs require tens of picometer wavefront error stability. This extreme stability is incompatible with the vibration from existing cryocoolers. Since ultra-low vibration cooling is a necessary prerequisite to using cryogenic single photon detectors on a LUVOIR, we briefly describe a few preliminary concepts for achieving it in Sec.~\ref{sec:cooling}. Although vibration will undoubtedly present challenges in starshade missions too, we believe that the coronagraphic LUVOIR represents a challenging ``worst case'' for cooler design studies.

In the interest of brevity, we have limited discussion to a fairly short list of detector technologies that are either already under development, or that we view as particularly promising. One could easily add other technologies to those that are discussed. For example, scientific CMOS arrays have a wide consumer base and potentially provide sub-electron read noise with better radiation tolerance than CCDs because no charge transfer is required. Superconducting nanowire single photon detectors (SNSPD) may provide another route to cryogenic single photon detectors that, while not energy resolving, would still promise essentially noiseless detection. As the field matures, it may be desirable to revisit these and other technologies. Although the need for essentially noiseless detection is clear, no existing technology currently fulfills all of the needs.

\section{Why better detectors are needed}\label{sec:justification}

Spectroscopic biosignature characterization places some particularly challenging demands on VISIR detector systems. Many of these derive from the extraordinarily low flux levels (Sec.~\ref{sec:starved}). In the case of superconducting detectors, achieving sufficient energy resolution and photon coupling efficiency are also challenges.\footnote{This paper's focus is on VISIR detectors for biosignature characterization. For other science programs, a general purpose LUVOIR would benefit from better detectors across its full $\rm 90~nm-2.5~\mu m$ ``stretch'' wavelength range, including the UV. We refer the interested reader to Bolcar~\etal,\cite{Bolcar:2016to} for a discussion of some of these other detector needs.}

\subsection{Biosignature Characterization}\label{sec:biosig}

\begin{figure}[t]
\begin{center}
\includegraphics[width=5in]{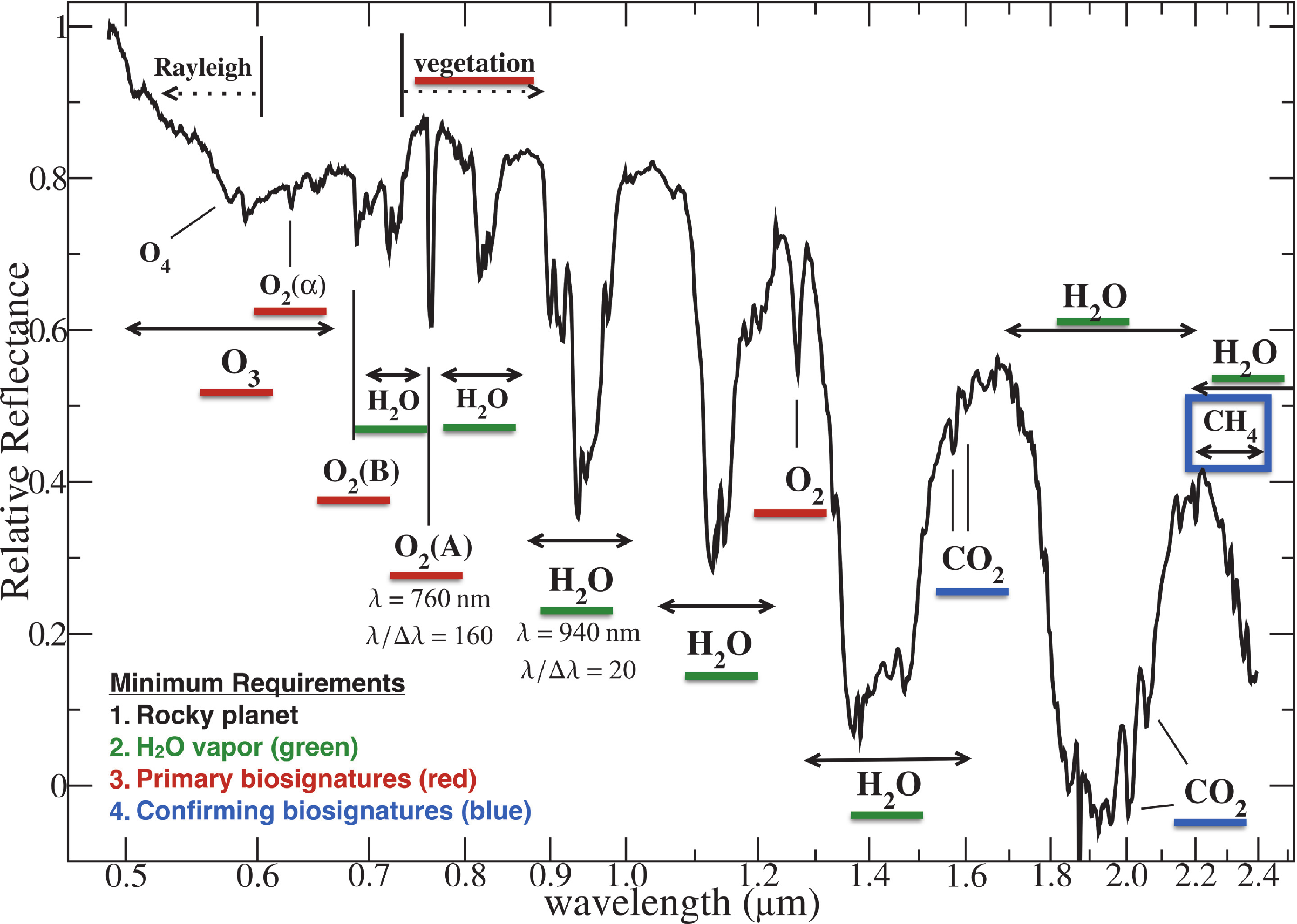}
\caption{\label{fig:biosignatures}Biosignatures are atmospheric spectral features that are thought to be necessary for life or than can be caused by it. Once a planet is known to be rocky and in the habitable zone, water is necessary to support life as we know it. Once water is present, biological processes can make \oxygen and \ozone, although other origins (\eg photo-disassociation of \water vapor) are also possible. A confirming biosignature, such as \methane, is helpful to rule out false positives. The \methane feature at 2.3~$\mu$m is unfortunately blended with \water vapor. If \methane is not useful, then detecting several biosignatures can be used to increase the statistical weight of findings. The vegetation red edge (VRE) is caused by chlorophyll from plants. Chlorophyll is expected to be difficult to detect in exoplanets.\cite{2014PNAS..11113278B} This figure overlays labels on Turnbull \etal's ``earthshine'' spectrum (see Ref.~\citenum{2006ApJ...644..551T}, Fig.~7). The resolutions that are given for \oxygen at 760~nm and \water at 940~nm are based on our least squares fits of gaussian profiles to Turnbull~\etal's data.}
\end{center}
\end{figure}

For the most likely potential exoEarths, biosignature characterization will be used to study spectral features that are thought to correlate with biological activity. Fig.~\ref{fig:biosignatures} shows how the earth would appear if it were to be seen as an exoplanet. To make this spectrum, Turnbull~\etal\cite{2006ApJ...644..551T} observed the night side of the moon and solved for the earth's contribution as it would appear to a distant observer. We define a likely life ``detection'' as consisting of; (1) a rocky planet, (2) with water vapor, (3) and a primary biosignature, and (4) a confirming biosignature to rule out false positives.

Lacking a confirming biosignature, one could attempt to increase the statistical significance of a result by resolving the temporal dependence of a feature. Arguments for a biological source could be strengthened by placing a detection in a more comprehensive geological and astrophysical context by measuring other atmospheric gases including \carbonmonoxide,  \cotwo, $\rm O_4\xspace$, and characterizing the host star's energy distribution.

Among confirming biosignatures, \methane is particularly important because it is difficult to simultaneously maintain significant concentrations of \oxygen, \ozone, and \methane. Non-equilibrium concentrations are most straightforwardly explained by biological processes. The \methane feature at 2.32~$\mu$m is unfortunately blended with \water. There is stronger \methane feature between 3~$\mu$m and 3.5~$\mu$m, and a still stronger feature at about $8~\mu\rm m$.\footnote{These longer wavelength lines would present other challenges, including reduced angular resolution (for a fixed aperture) and potentially increased thermal background.} The spectrum shows a few other features, notably ${\rm CO_2}$ and ${\rm O_4}$. Although these features do not provide as much information as the primary and secondary biosignatures, they can still be useful, especially when no confirming biosignature is available.\cite{Schwieterman:2016kz} For a thorough discussion of several false positive mechanisms and their spectral signatures, the interested reader is referred to Schwieterman~\etal\cite{Schwieterman:2016kz}

Fig.~\ref{fig:biosignatures} omits one important near-UV biosignature. 
There is a strong \ozone bandhead at $260-350~\rm nm$. This bandhead is so strong and wide that it can potentially be characterized by imaging in a pair of filters. Because our focus here is on spectroscopic biosignature characterization, we defer discussion of these (potentially imaging) near-UV detectors to a future publication. Bolcar~\etal\cite{Bolcar:2016to} discusses detectors for this application in slightly more detail (see especially his Tab.~6).

Finally, the earth's atmosphere has not always been as it is today, and it is conceivable that other atmospheres may harbor life.\cite{Wagner:2011wq} For these reasons, we should be open to the possibility of having to characterize several spectral features in order to understand how likely an exoplanet is to harbor life. Having the best detectors possible will maximize our chances of success.

\subsection{Required Spectral Resolution}\label{sec:resolution}

Several authors have studied the spectral resolution requirements for biosignature characterization.\cite{2002AsBio...2..153D,Kaltenegger:2007cg,2014PNAS..11113278B} Their recommendations vary depending upon the spectral features of interest and the model assumptions. As can be seen from Fig.~\ref{fig:biosignatures}, important features include \water, \oxygen, \ozone, and \methane. All have absorption features in the VISIR and are important to terrestrial life. Consistent with Brandt and Spiegel (2014),\cite{2014PNAS..11113278B} we have adopted \oxygen as a challenging, but probably still achievable biosignature upon which to base our VISIR detector requirements because it is the narrowest feature on this list. An instrument that can characterize \oxygen can also characterize \ozone and \water, and potentially other features including \methane and \cotwo under the right conditions (see \eg Ref.~\citenum{Kaltenegger:2007cg}).

With regard to the spectral resolution value, Des Marais~\etal (2002)\cite{2002AsBio...2..153D} reported that $R=69-72$ was well matched to \oxygen at 760~nm and 1.27~$\mu$m (see their Tab.~1). This was based on a theoretical model that included the Earth's current atmospheric temperature structure, but that allowed for different chemical abundances. More recent studies have recommended that somewhat higher resolution is desirable for \oxygen. For example, when Kaltenegger, Traub, and Jucks (2007; KTJ)\cite{Kaltenegger:2007cg} modeled the evolution of the expected spectra of the Earth and its biosignatures over geological timescales and found that $R=125-136$ was optimal for observing \oxygen in the visible. They furthermore concluded that higher resolution, $R=165-244$, would be desirable to observe \oxygen in the near-IR. More recently, Brandt and Spiegel (2014)\cite{2014PNAS..11113278B} recommended that $R=150$ was probably adequate for \oxygen in the VISIR. This is also consistent with the \oxygen line width that we measured from Turnbull~\etal's ``earthshine'' spectrum (Fig.~\ref{fig:biosignatures}). Our $R>100$ requirement represents a working compromise between the still evolving scientific understanding and the practicalities of developing flight hardware. As the field matures, it may be necessary to revisit this requirement.

If \oxygen were not required, then lower resolution could be tolerable depending on the scientific objectives. For example, KTJ found that $R=8-11$ would be sufficient to characterize \water in the VISIR throughout Earth's evolution, and that $R=4-5$  would be sufficient for \ozone during specific epochs. However, KTJ's recommended range was wide, and they concluded by recommending $R=8-325$ to detect \water, \cotwo, \oxygen, and \methane in the VISIR. To better understand the full biosignature trade space with regard to spectral resolution, we refer the interested reader to Refs.~\citenum{2002AsBio...2..153D,Kaltenegger:2007cg,2014PNAS..11113278B}.

\subsection{Photon starved science}\label{sec:starved}

Once the light from the host star has been suppressed, the remaining light from the exoplanet and its zodiacal cloud will be feeble at best. To put the photon arrival rate into better perspective, consider a simple toy model consisting of: (1) a perfect coronagraph, (2) a 25\% efficient integral field unit (IFU) spectrograph, (3) a $\lambda=550$~nm observing wavelength, (4) pixel size $=0.7\times 1.22\lambda/D$, (5) $R=150$, and (6) a background that is $3\times$ the earth's zodiacal light. With these assumptions, the background count rate is $<0.001~\textrm{cts}~s^{-1}~\textrm{pix}^{-1}$. More sophisticated models that include the effects of imperfect coronagraphs and simulated exoEarths reach the same conclusion: biosignature characterization is extremely photon starved.\cite{Stark:2015er}

The preceding calculation assumed a non energy resolving detector behind a conventional IFU spectrograph. Use of an energy-resolving single photon detector would eliminate the need for spectrograph optics and increase the count rate per pixel by about a factor of $100\times$ (see Appendix~\ref{sec:count-rate} for the derivation). Under these conditions, the count rate would be about $0.1~\textrm{cts}~ s^{-1}$ per energy resolving pixel.

\subsection{Strawman Detector Needs}

Tab.~\ref{tab:req} shows the detector requirements that were used for NASA's recently completed ATLAST study.\cite{Bolcar:2016to} We adopt these as the basis for further discussion. We are aware of additional desirable characteristics. For example, a starshade-based habitable exoplanet mission might benefit from response further into the IR than the ATLAST team considered. We have tried to note these other needs as they come up.

Taken collectively, the ``requirements'' of Tab.~\ref{tab:req} enable characterization of a few dozen exoEarth candidates during an approximately five year LUVOIR-like mission. Stark~\etal\cite{Stark:2015er} provides a good overview of the mission yield modeling.

\begin{table}[t]
\caption{Strawman ATLAST Detector Needs} 
\label{tab:req}
\begin{center}
\includegraphics[width=0.66\textwidth]{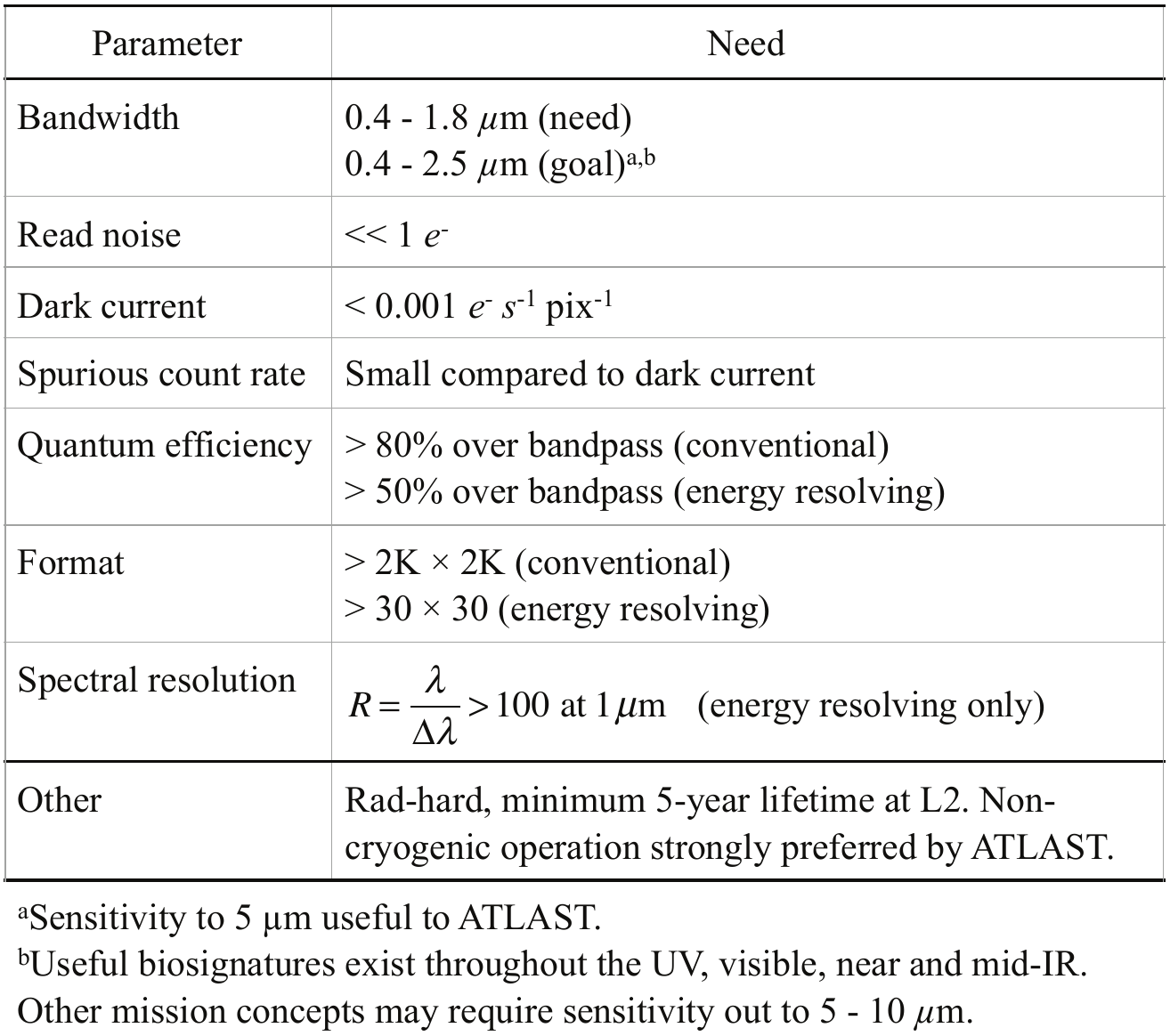}
\end{center}
\end{table} 

The QE requirements merit further discussion. To achieve reasonable exoEarth yields, all mission concepts that we are aware of assume no QE penalty compared to today's best EMCCDs and HgCdTe IR arrays.\cite{Bolcar:2015jh,2015arXiv150704779D,Stark:2015er}. They also assume megapixel class detector arrays paired with conventional spectrographs. If an energy resolving detector were to be used, then the spectrograph optics could be greatly simplified and the optical throughput would go up. For this reason, lower QE can be tolerated with a single photon detector than with a conventional detector.

\section{Improving today's state of the art}\label{sec:btb}

The most mature VISIR detector candidates are semiconductor based. These include silicon EMCCDs for the visible and HgCdTe photodiode and avalanche photodiode (APD) arrays for the VISIR. EMCCDs, HgCdTe hybrids, and HgCdTe APD arrays are attractive because of their comparative maturity, low risk, and the possibility that their performance might be ``good enough'' for biosignature characterization, even if they do not function as single photon detectors.

For use in space, radiation tolerance is a major consideration. Existing e2v EMCCDs may not be sufficiently radiation tolerant (``rad-hard'') for future biosignature characterization missions. Current generation e2v EMCCDs were designed for use on the ground, and are based on an n-channel CCD architecture for which phosphorus is the dopant. We discuss how the radiation tolerance of EMCCDs could be enhanced in Sec.~\ref{sec:emccd}.

Teledyne's HxRG photodiode arrays, like JWST's H2RGs and the closely related WFIRST H4RG-10s,\footnote{WFIRST's H4RG-10s still require radiation testing. However, based upon our knowledge of the components, we expect the radiation tolerance to be similar to that of JWST's H2RGs although the degradation rate in pixels per year may differ on account of the smaller pixel size.} are radiation tolerant. JWST testing has shown that H2RGs experience graceful degradation in pixel operability, whereby approximately 2-3\% of the pixels per year will degrade to the extent that they no longer exhibit full science performance. Although the affected pixels no longer meet full flight specification, they can still be useful for many things, and in any case only a small percentage of pixels are affected at the end of JWST's nominal five year mission. The radiation tolerance of HgCdTe APD arrays is to be determined, but in any case the failure modes that are seen in n-channel CCDs do not apply.

\subsection{Better EMCCDs}\label{sec:emccd}

e2v EMCCDs are widely regarded as the most mature detector technology for visible wavelength biosignature characterization today. For this reason, $\rm 1K\times 1K$~pixel e2v CCD201s have been selected for the WFIRST coronagraph's imaging camera and integral field spectrograph. Harding and Demers \etal\cite{Harding:2016bc} describe the extensive trade study that led to this selection. When new and not degraded by radiation, EMCCDs are close to meeting the needs for biosignature characterization.

Unfortunately, a major concern with the current e2v EMCCD design for biosignature characterization on LUVOIR or a Habitable Exoplanet Imaging Mission is radiation induced performance degradation. This may include decreased charge transfer efficiency, increased clock induced charge (CIC), and decreased pixel operability. Moreover, the sub-electron read noise that EMCCDs enable has the potential to reveal other damage that is ordinarily buried in the $2-3~e^-$ read noise of conventional CCD systems. Although ongoing work at JPL should retire these concerns for WFIRST,\cite{Harding:2016bc} the demands of future biosignature characterization missions will be more challenging. For future missions, it would be wise to apply e2v's known radiation hardening fabrication processes to EMCCDs and to explore other rad-hard detector concepts.

Existing e2v EMCCDs use gate oxide designs that were intended to maximize manufacturing yields for less demanding ground-based applications. They trade radiation tolerance in exchange for lower manufacturing cost. The oxides are thicker and of a different composition to those that are used in radiation hardened CCDs. Radiation hardened oxides can reduce the flat-band voltage shift from $\rm \sim 100-200~mV~krad^{-1}~(Si)$ in standard devices to $\rm 6~ mV~krad^{-1}~(Si)$ in devices fabricated using radiation-hardened oxides.\cite{Burt:2009ct}

It would also be desirable to explore design enhancements for reducing CIC. CIC is strongly dependent upon clock amplitude. In the CCD201, there are boron implants beneath two of the phases that build in the electric fields that are needed for inverted operation. These inherently represent a compromise. Making them stronger increases well depth, but it comes at the expense of CIC when reading the CCD out on account of the higher voltages that are required. For biosignature characterization, it could be worthwhile to explore implant designs that aim to trade well depth for improved CIC.

Thinner and different oxide designs may also be beneficial for reducing CIC, which Janesick attributes to hole detrapping near the silicon/oxide interface.\cite{Janesick:2001wb} Burt~\etal\cite{Burt:2009ct} attribute some of the performance degradation that is seen with radiation dose to depassivation of the silicon surface under the oxides, and moreover suggests that thinner oxide layers should result in less depassivation. If true, one might reasonably expect to see less CIC degradation in parts that use thin vs thick oxide layers.

\subsection{Ultimate limits of HgCdTe photodiode arrays}\label{sec:hgcdte_limits}

\HgCdTeL is today's most mature material for astronomical near-IR instruments. By adjusting the mole fraction of cadmium, $x$, it is possible to tune the cutoff wavelength from about 1.7~$\mu$m out to $5-10~\mu\rm m$ while still achieving performance that enables low background space astronomy. HgCdTe arrays have substantial heritage for NASA astronomy. The Hubble Space Telescope has operated both NICMOS and H1R HgCdTe arrays. Teledyne H2RGs are used by all of JWST's near-IR instruments and by Euclid. Teledyne H4RG-10s are planned for WFIRST. The read noise floor of existing HgCdTe photodiode arrays is a few electrons rms per pixel. When cooled sufficiently, the dark current of today's $\rm 2.5~\mu m$ cutoff flight grade HgCdTe arrays already achieves the $<0.001~e^-~s^{-1}~\rm pix^{-1}$ that is needed for biosignature characterization.

The source-follower-per-detector architecture of Teledyne's HxRG arrays has been used since the late 1980s. To achieve significant improvement in noise, it is necessary to understand  exactly where in HxRG arrays read noise and dark current originate and why. Studies that aim to separate the contributions of the photodiodes, resistive interconnects, ROIC source-followers, and SIDECAR controllers would be beneficial.

For example, if it were to be found that noise in the resistive interconnects were to be important, than further work aimed at reducing the interconnect resistance and/or lower operating temperature could be beneficial. On the other hand, if noise from the pixel source-follower were to be important, than further refinement of this circuit might be justified. The first step is careful characterization of existing HxRG detectors, (\ie JWST and WFIRST spares) aimed at building an itemized noise budget and understanding how environmental parameters like operating temperature affect performance.

Another area where improvement might be possible is persistence. Persistence, or latent charge, is charge that accumulates and is trapped during an exposure only to be released as an undesirable ghost signal in a subsequent exposure. Persistence is modulated by charge traps, or electrically active defect states in the HgCdTe. Design and process improvements that aim to reduce the defect density, or that aim to build in electric fields that repel charges from areas of high defect density, could be beneficial for reducing persistence.

\subsection{HgCdTe APD arrays}\label{sec:apd_arrays}

HgCdTe APD arrays are a promising technology that initially entered astronomy for comparatively high background applications including adaptive optics and interferometry\cite{Finger:2012dv} and wavefront sensing and fringe tracking\cite{Finger:2014we}. More recently, they have been used at the telescope to provide diffraction-limited imagine via the ``lucky imaging'' technique.\cite{2014SPIE.9154E..19A} Although HgCdTe APD arrays have been made by DRS, Raytheon, and Teledyne; those made by Selex in the UK are the focus of most attention in astronomy now.

A group at the University of Hawaii has been evaluationg Selex SAPHIRA for applications including low background astronomy.\cite{2014SPIE.9154E..19A} With appropriately optimized process, the HgCdTe itself is potentially capable of the same QE performance as the JWST arrays.\footnote{JWST's H2RG's achieve $\rm QE>70\%$ from $\rm 0.6-1~\mu m$ and $\rm QE>80\%$ from $\rm 1-2.5~\mu m$.\cite{Rauscher:2014wk}} Moreover, because gain is built into the pixels before the first amplifier, they promise photon counting and potentially even single photon detection if ``dark current'' can be reduced to acceptable levels.

``Dark current'' is the most significant obstacle to using Selex APD arrays for ultra-low background astronomy today. The $\sim 10-20~e^-~s^{-1}~\textrm{pixel}^{-1}$ gain corrected ``dark current'' that has been reported\cite{2014SPIE.9154E..19A} is almost certainly dominated by glow from the ROIC. The ROIC that is used in current devices was not optimized for ultra-low background, or even low background astronomy. Work continues at the University of Hawaii to try to disentangle ROIC glow from more fundamental leakage currents in current generation APD arrays. On the longer term, work is also underway aimed at optimizing the ROIC design.

Although HgCdTe APD arrays hold out the promise of read noise below that which can be achieved using conventional photodiode; like conventional photodiodes there will ultimately be a leakage current noise floor that is determined by thermally activated defect states in the HgCdTe. However, it is likely that today's performance is still far from that floor, and more work is needed to better understand the full potential of HgCdTe APD arrays for ultra-low background astronomy in the context of missions like LUVOIR.

\section{Maturing energy resolving Single Photon Detectors}\label{sec:spd}

Today's EMCCDs, HgCdTe hybrids, and HgCdTe APD arrays are not single photon detectors in the context of biosignature characterization. All would add significant noise and thereby reduce mission exoEarth yields below what could be achieved with a noiseless detector. On the other hand, superconducting MKID and TES arrays already function as single photon detectors today.

The use of these superconducting detectors by LUVOIR is contingent upon the development of ultra-low vibration cooling (Sec.~\ref{sec:cooling}). However, even if superconducting detectors are found to be impractical for LUVOIR, their nearly quantum limited performance could still be very attractive for a starshade-based Habitable Exoplanet Imaging Mission.

\subsection{Introduction to superconducting proportional detectors}\label{sec:cryo_intro}

Although proportional detection of photons is not widely used for VISIR astronomy today, it has a long history in X-ray astronomy where gas proportional counters and Ge and Si diodes have long been standard detectors. These charge-based detectors are proportional in the sense that their response to light is proportional to photon energy.  Although they provide an easy to measure signal, they suffer from noise sources that make high resolution spectroscopy impossible.  In both cases, only part of the signal goes into ionization, so there is an unavoidable partition noise. This problem can be addressed in two ways;  either to move to a much smaller gap in the detection system, or to collect the energy into a gapless system, such as a thermal distribution of phonons and/or electrons.

The small gap solution leads us to superconducting detectors, superconducting tunnel junctions or kinetic inductance detectors, which measure the quasiparticle excitations produced by the absorption of a photon. The superconducting gap is about a factor of 1000 smaller than typical semiconductor bandgaps, so the expected energy resolution at 6~keV improves from about 130~eV for Si diodes to 2.7~eV for Nb superconducting detectors.\cite{1982NucIM.196..275K}  An alternative approach is to let all of the deposited energy thermalize, and then measure the temperature of the system. This is a quasi-equilibrium system, and it is a robust measurement technique. For X-ray detection the uncertainty of energy measurements easily reaches limits set by the thermodynamics of the system, and with proper design, is never limited by the ability to thermalize the photon energy.  

Thermal sensors simultaneously detect individual photons and use the thermal signal to measure photon energy. For the designs discussed here, the minimum photon energy is well separated from the system noise, so the probability of dark events are near zero, and the “read noise” manifests itself as the limit to the energy resolution of the system.  In the following sections, we will discuss the performance of cryogenic proportional photon detectors, TES microcalorimeters (Sec.~\ref{sec:tes}) and MKIDs (Sec.~\ref{sec:mkid}) as VISIR spectrometers.

\subsection{Transition-edge sensor (TES) microcalorimeter arrays}\label{sec:tes}

In a microcalorimeter, the energy of an absorbed photon is determined from the temperature rise of the detector.  The energy resolution of such a detector is set by thermodynamic noise sources in the detector and amplifier noise. The microcalorimeter concept and performance limits are presented by Moseley, Mather, and McCammon (1984)\cite{1984JAP....56.1257M} and Irwin~\etal (1995)\cite{Irwin:1996fe}.   Tutorials articles on the principles of operation and optimization of microcalorimeters and  superconducting TESs provide a detailed discussion of these devices in the linear regime.\cite{Enss:2008jz,2005cpd..book...63I} Since these devices work near equilibrium, it is generally possible to design detectors that operate very near these fundamental limits. \cite{2005cpd..book...63I,2008JLTP..151..406I,2008JLTP..151..400B}

When a photon is absorbed by the detector, the temperature rises on a short time scale, of order the sound crossing time.  The output signal will rise on a time scale set by the electronic time constant of the detector/amplifier combination.   The detector will rise to a maximum temperature $\delta T \sim \delta E / C$, where $C$ is the lumped heat capacity and $\delta E$  deposited energy. A simple microcalorimeter, modeled as a lumped heat capacity and thermal conductance, $G$, will have a single-pole response, with a time constant $\tau = C/G$.  Under bias, the response is sped up by electrothermal feedback to $\tau_e = \tau /(1+ (P \alpha/G T_0))$, where $P$ is the Joule power in the detector and $T_0$ the detector temperature.

The energy resolution of a microcalorimeter scales as  $\sqrt{k_B T^2 C/\alpha}$, where $T$ is the temperature, $k_B$ is the Boltzmann constant, and $\alpha=d \log R/d \log T =(T/R) (dR/dT)$ is a unitless measure of the sensor's resistance sensitivity to temperature.\cite{1984JAP....56.1257M} Elsewhere in this article, $R$ represents spectral resolution. Here $R$ represents resistance. This resolution limit assumes we remain in the linear response range, and uses an optimal detection filter based on the system noise and signal shape. The thermodynamic performance does not depend on the choice of $G$; it can be used to minimize other non-optimal effects in the detector, such as slow thermalization.

A typical design for such a detector would set the heat capacity for a given value of $\alpha$ to allow saturation to begin just above the highest energy of interest.  Having chosen this value of $C/\alpha$, the resolving power of the linear system can be improved only by lowering the operating temperature.

The first demonstration of a VISIR microcalorimeter was presented in 1998 by Cabrera~\etal\cite{1998ApPhL..73..735C} In this work, they built a tungsten TES microcalorimeter $18~\mu\rm m$ square with a transition temperature of $100~\rm mK$. The thermal conductance of these detectors was set by their internal electron-phonon coupling, so they were deposited on a substrate, requiring no additional thermal isolation. These devices were operated in the VISIR spectral region, and provided an energy resolution of $0.15~\rm eV$. This was significantly in excess of the na{\"i}ve prediction, but a more complete analysis that we have done, including the current dependence of the resistance $(\beta)$ and athermal phonon loss to the substrate, can account for most of the excess noise. Both of these terms can be significantly reduced by design optimization.

In considering these detectors as candidates for the characterization of exoplanet atmospheres, we need to explore the paths to achieving the resolving power $R>100$ that is required. This optimization must include the efficient coupling of the detectors to the optical photons as well as providing the required energy resolution. The coupling design may be as simple as absorption by a matched film, as in the case of Cabrera~\etal,\cite{1998ApPhL..73..735C} or may require an antenna structure to couple to optically small detectors. We explore three optimizations (Tab.~\ref{tab:three_optimizations}) and compare the system resources required in each case.  For the natural time constant $\tau$ row in Tab.~\ref{tab:three_optimizations}  we have assumed a tungsten TES with $C$ and $G$ determined by its electron system and electron-phonon coupling respectively.

\begin{table}[t]
\begin{center}
\caption{Three optimizations for TES microcalorimeters$^a$}\label{tab:three_optimizations}
\includegraphics[width=.8\textwidth]{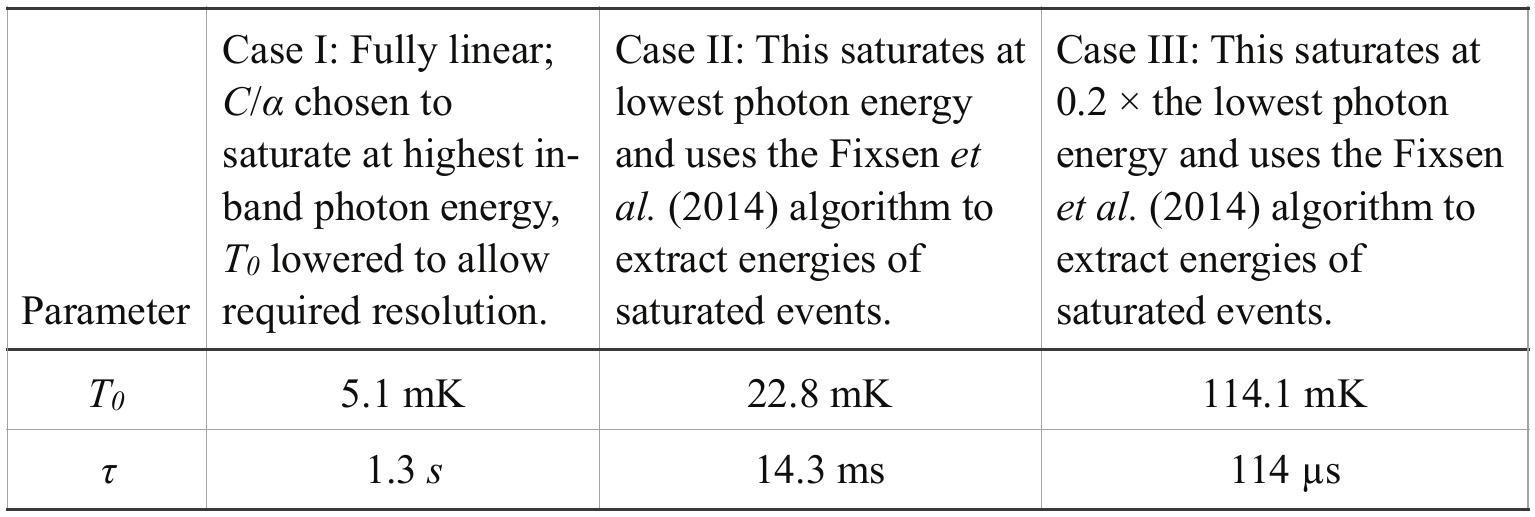}
\parbox{.8\textwidth}{$^a$Fixsen~\etal (2014) appears here as Ref.~\citenum{Fixsen:2014jl}.}
\end{center}
\end{table}

These designs provide possible paths to practical detectors for biosignature characterization. More specifically, they potentially provide the noiseless detection of photons with intrinsic resolving power $R\sim 100$.  By operating in the nonlinear regime, we should be able to reach the required performance at $T\sim 50~\rm mK$, temperatures already demonstrated by ADRs designed for space (see Sec.~\ref{sec:cooling}). We furthermore believe that the near equilibrium operation should  allow us to approach closely the predicted performance.  

\subsection{Microwave kinetic inductance devices (MKID)}\label{sec:mkid}

This section briefly reviews MKIDs in general, and in particular describes what makes a VISIR MKID.  We consider the state of the art in the context of MKID device physics, and paths for improving energy resolution, quantum efficiency, and pixel count to achieve the performance goals specified in Tab.~\ref{tab:req}.

An MKID detects absorption of photons in a superconductor by a change in kinetic inductance.  A current in a superconductor stores energy both in magnetic field and in kinetic energy of the charge carriers (Cooper pairs).  The former corresponds to the geometric inductance of normal conductors, while the latter represents additional, kinetic inductance.  For a photon to be absorbed, its energy must exceed twice the superconducting gap energy $\Delta$ (the binding energy per electron in Cooper pairs) in order to create unpaired electrons (quasiparticles).  A reduced number of Cooper pairs requires the remaining Cooper pairs to move faster to transport the same current, thereby increasing kinetic inductance.  While typical MKID material contains on the order of one to ten million Cooper pairs per cubic micron in its superconducting state, the temporary destruction of even one Cooper pair is possible to detect if a small volume inductor is combined with a (distributed or lumped) capacitor to form a low-loss superconducting resonator with resonance frequency $ \approx $ 1 GHz.  An exceedingly small change in inductance can be sensed as a shift in a resonance frequency, measured with a commercially available microwave amplifier.  

In typical MKID materials, the superconducting transition temperature is of order 1 K, for which the energy gap is 0.15 meV, and the minimum frequency for photon absorption is 74 GHz ($\lambda \approx 4  \; \rm mm$).  Absorption of a single VISIR photon creates a large number of quasiparticles $ (\approx 10^{3} - 10^{4}) $ in proportion to the photon energy.  The MKID operating temperature and inductor volume can be made small enough so that the number of thermally generated quasiparticles is nil.  MKIDs are then energy resolving detectors with zero dark count rate.

In addition to high sensitivity, MKIDs have a natural means of multiplexing:  large numbers of superconducting resonators, tuned to slightly difference resonance frequencies, can be connected in parallel by injecting a comb of microwave carrier waves on one transmission line, and read out by one microwave amplifier.  Systems have been demonstrated for simultaneous readout of up to 4000 MKIDs.\cite{vanRantwijk:2015un}

Owing to sensitivity at long wavelengths, and the ability to multiplex tens of thousands of detectors, there has been a wealth of work directed at the potential of MKIDs for sensitive detection of FIR to mm-wave radiation.  In pushing towards single photon sensitivity in this spectral region, the fundamental noise contributions in MKIDs have been studied extensively, and there has been much innovation in optical coupling schemes (see \eg Refs.~\citenum{Zmuidzinas:2012kh,2008PhDT.......222G}).  Here we discuss a few of the differences in approach relevant at the higher photon energies in VISIR MKIDs.

The MKID signal is a change in the amplitude and phase of a microwave carrier tone propagating on a transmission line weakly coupled to the MKID resonator via some auxiliary coupling capacitance or inductance.  The amplitude and phase shift is proportional to the number of quasiparticles $ N $ produced by absorption of a photon, $N = \eta h \nu / \Delta $, where $ \nu $ is the photon frequency and $ \eta $ is the efficiency.  MKIDs have two fundamental sources of noise: (1) quasiparticle generation-recombination (G-R) noise, and (2) microwave amplifier noise.  Quasiparticles are generated not only by optical absorption, but also by thermal fluctuations (e.g. thermal phonons).  The quasiparticles fluctuate in number as they are constantly being generated, and then recombining into Cooper pairs at a rate, specific to the material, that increases in proportion to the density of quasiparticles.  As the temperature or optical load increases, the increasing population of quasiparticles both adds noise\cite{Yates:2011jy} and increases microwave dissipation (lowering the quality factor $ Q $).  Maintaining a sufficiently high $ Q $ is required both because it lessens the importance of amplifier white noise, but also because the number of detectors that can be multiplexed within the amplifier bandwidth is proportional to $ Q $ (or, more precisely, to the effective $Q$, $Q_{eff} = \pi \tau f_{c} $, where $f_{c}$ is the microwave carrier frequency and $\tau^{-1}$ is the optimally-filtered pulse detection bandwidth determined by the $Q$ and the noise sources).  

\begin{figure}[t]
\begin{center}
\includegraphics[width=\textwidth]{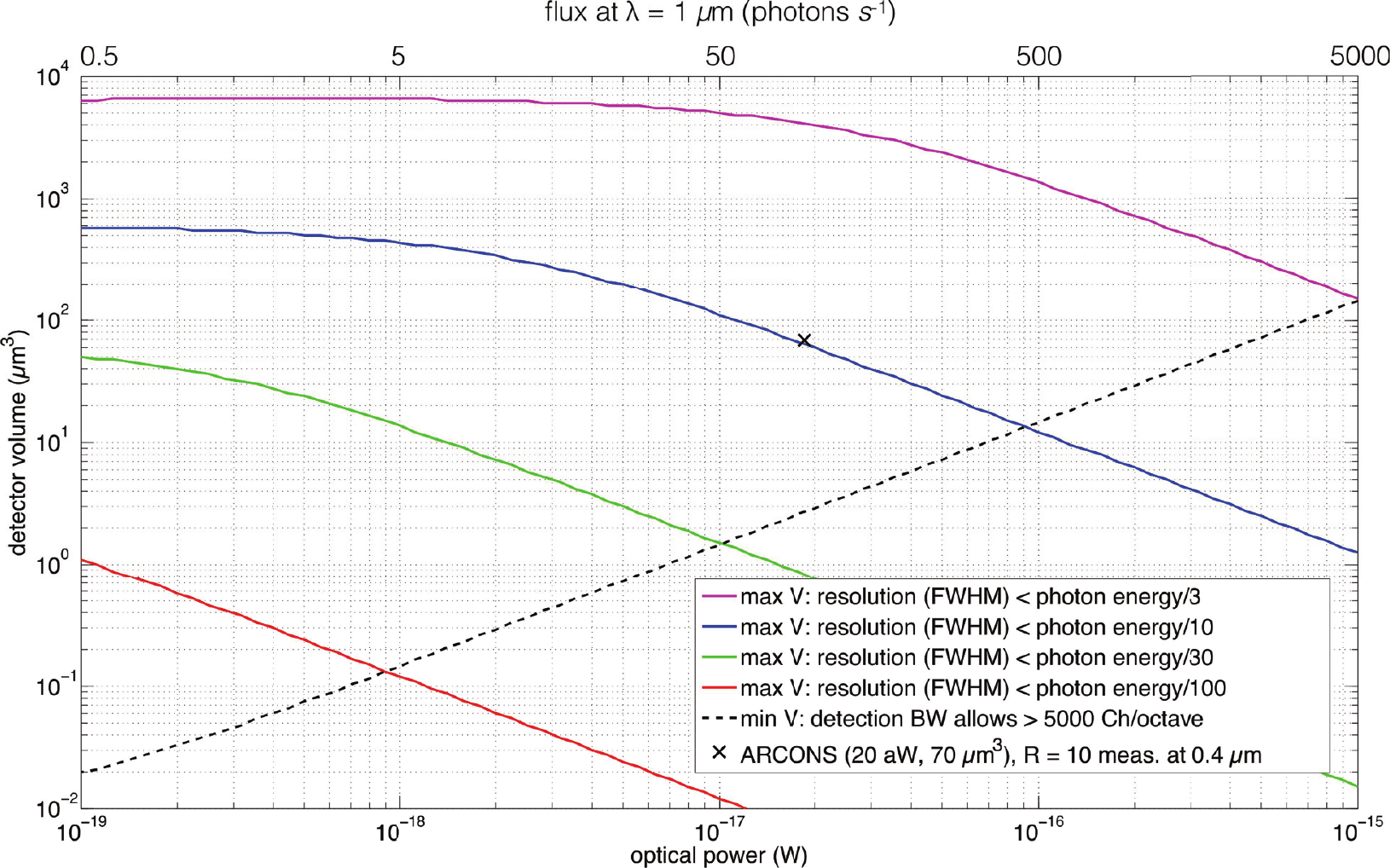}
\caption{\label{fig:ARCONS_R10_now}Example of current state of the art of VISIR MKID design: Various requirements for photon counting performance are shown as curves bounding allowed regions of the detector volume – optical power phase space.  Point marked (x) shows ARCONS existing VISIR MKIDs with resolving power 10 for optical load power 20 aW and detector inductor volume 70 cubic microns. The models were run in units of Watts. We convert this to approximate photons per second at $\lambda=1~\mu$m on the top axis for presentation purposes.}
\end{center}
\end{figure}

Other, non-fundamental sources of MKID noise exist (e.g. fluctuations in resonator capacitance from two-level-systems (TLS) in disordered native oxides or other dielectrics present), and are the subject of active research.  However, G-R and amplifier noise, and the associated effects on $ Q $, determine some general aspects of MKID optimization.  For photon-counting capability, there is a maximum inductor volume that gives the desired energy resolution and a detector speed faster than the photon arrival rate.  There is also a minimum volume that keeps the $ Q $ under optical loading high enough for practical multiplexing.  Fig.~\ref{fig:ARCONS_R10_now} illustrates these design constraints applied to MKIDs, made from TiN, with parameters similar to those of the state-of-the-art ARCONS VISIR MKID detectors.\cite{2013PASP..125.1348M}

In Fig.~\ref{fig:ARCONS_R10_now}, the x-axis is chosen to be the mean optical power absorbed.  At high powers, the quasiparticles produced from one photon are still present in significant numbers when the next photon arrives.  At low powers, the MKID recovers to nearly zero quasiparticle number between photons; in Fig.~\ref{fig:ARCONS_R10_now}, this corresponds to the curves of fixed energy resolution flattening and becoming independent of optical power (photon rate) at sufficiently low power levels.  The maximum photon rate for this regime is not as fast as one would expect given that the initial time constant for decay of the quasiparticle number is typically quite fast ($< 100 \mu$s) because the decay is not exponential.  The rate of recombination events is proportional to the square of the number of quasiparticles present, yielding a mean number of quasiparticles at time $t$ given by $N(t) = (N(0)^{-1} + tR_{qp}/V)^{-1}$, where $R_{qp}$ is the recombination rate constant for the material, and $V$ is the detector volume.  While we have carried out Monte Carlo simulations of the time-domain MKID waveforms for random photon arrival streams, the simpler approximate treatment represented by Fig.~\ref{fig:ARCONS_R10_now} gives energy resolution and bandwidth results sufficiently accurately for this discussion.

For long-wavelength applications, one goal is generally for MKID sensitivity to reach the photon-noise background-limited Noise Equivalent Power (NEP), or to count mm-wave photons with resolving power $R > 1$.  However, the goal for VISIR MKIDS is much higher $R$.  ARCONS MKIDs achieve $R = 10$ for $\lambda = 0.4~\mu\rm m$.  While Fig.~\ref{fig:ARCONS_R10_now} indicates the observed resolving power is about what is expected, more detailed detector models than used here in making Fig.~\ref{fig:ARCONS_R10_now} lead the ARCONS team to conclude that their resolving power is a factor of two less than expected due to a peculiarity of the microstructure of TiN films that gives a photon-absorption-position dependent responsivity.   Additionally, resolution is expected to improve when using parametric amplifiers (currently under development) to gain a large ($>10\times$) improvement in amplifier noise temperature compared with HEMT amplifiers. However, improving VISIR MKIDs to $R = 100$ faces a significant challenge in that there is statistical noise on the quasiparticle creation process.  Absorption of a photon well above the gap frequency initially forms a pair of high energy quasiparticles, which decay in energy by emitting phonons sufficiently energetic to break additional Cooper pairs.  In the cascade of quasiparticles and phonons produced, the fraction of the original photon energy lost to the substrate by phonons is variable.  Consequently the energy resolution is subject to the Fano statistics limit: $ R < \sqrt{\eta h \nu /(F\Delta)}/2.3458 $, where $\eta = 0.58$ and $F = 0.2$ is the Fano factor.\cite{1982NucIM.196..275K} The energy resolution will be additionally degraded if hot phonons escape to the substrate prior to completion of the cascade process to convert the photon energy into low energy photons and quasiparticles. \cite{Kozorezov2013} 

\begin{figure}[t]
\begin{center}
\includegraphics[width=\textwidth]{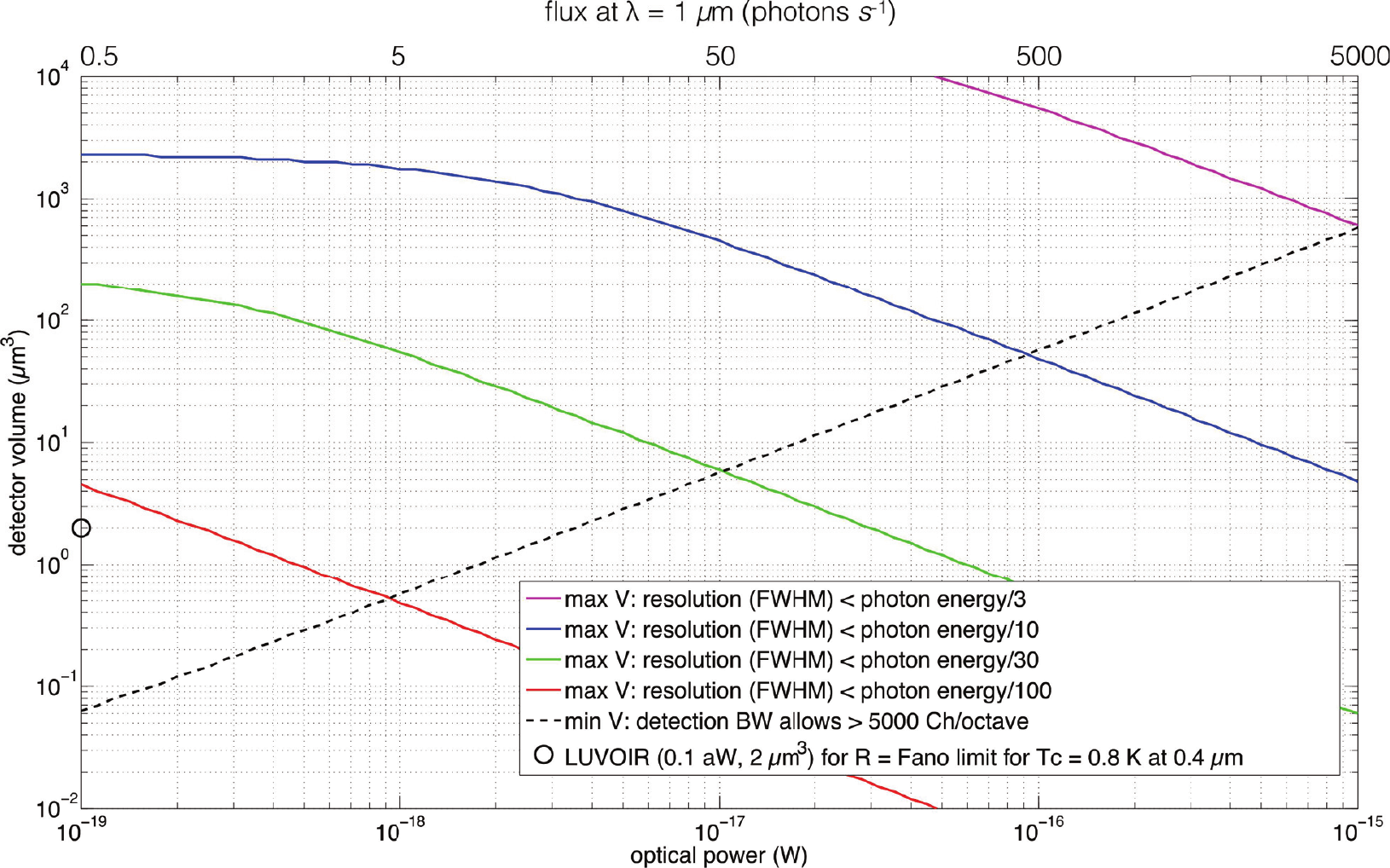}
\caption{\label{fig:ARCONS_R115}Phase space for achieving Fano-limited resolving power with VISIR MKID design with TiN film similar to ARCONS, but with a quantum-limited readout amplifier. Phase space for photon counting performance at 0.1 aW shrinks to a maximum inductor volume of 2 cubic microns. The models were run in units of Watts. We convert this to approximate photons per second at $\lambda=1~\mu$m on the top axis for presentation purposes.}
\end{center}
\end{figure}

Using $\Delta = 1.77 k_{B} T_{c}$, and say, $T_{c} = 1$~K, one finds the Fano limit is $R=42$ at $\lambda = 2.5~\mu\rm m$, and $R=104$ at $\lambda = 0.4~\mu\rm m$. Fig. ~\ref{fig:ARCONS_R115} shows the change in inductor volume to reach the Fano limit at $\lambda = 0.4~\mu\rm m$, assuming the switch to a parametric amplifier. For a LUVOIR optical power (0.05 aW for energy resolving detectors at 0.1 cps, instead of 20 aW for ARCONS ground based background at 75 cps), the Fano limit goal requires the detector volume be reduced from 70 to 2~$\mu\rm m^{3}$.  The volume reduction will start to push the MKID response into a non-linear regime, but yields the necessary sensitivity.  In this regime, the instantaneous shift in resonance frequency and $Q^{-1}$ during the pulse are both large, but evolve in tandem to give an approximately constant phase shift in the microwave carrier for some time.  The energy of the photon is then not just encoded in the pulse amplitude, but in the pulse duration.  (This saturation of the pulse amplitude is similar to the mode of TES operation advocated in the previous section.)  Keeping the ARCONS TiN thickness value of 60 nm, the required area for the inductor needs to shrink from $40\times 40~\mu\rm m^2$ square to $6\times 6~\mu\rm m^2$.  Optical coupling schemes are discussed below; however, while it may be possible to still couple optically to this inductor size, the Fano limit seems to preclude attaining the LUVOIR resolving power goal at the long wavelength end at $\lambda = 2.5~\mu\rm m$. However, depending upon the spectral features of interest, this may be acceptable since Fig.~\ref{fig:biosignatures} appears to suggest that high spectral resolution is most important for \oxygen and \ozone features in the visible. It would be helpful if atmospheric models could be used to better define the required spectral resolution as a function of wavelength throughout the VISIR.

One path to circumventing the Fano limit and hot phonon escape is to fabricate the MKID on a suspended membrane structure, potentially changing it from a pair-breaking detector to an equilibrium thermal detector in which the MKID serves only as a thermometer sensing the temperature rise of the membrane.  Such Thermal MKID (TKID) devices have already been made and tested for x-ray microcalorimetry.\cite{FirstResultsfromT:2016tb} Arrays of MKIDs on one common silicon membrane, rather than individual membranes, have also been demonstrated.\cite{Brown:2010gl}  The membrane design reduces the stochastic variation in detected energy at the cost of slowing the response time, but, for the low photon count rates in biosignature characterization, that may be acceptable.

\begin{figure}[t]
\begin{center}
\includegraphics[width=.7\textwidth]{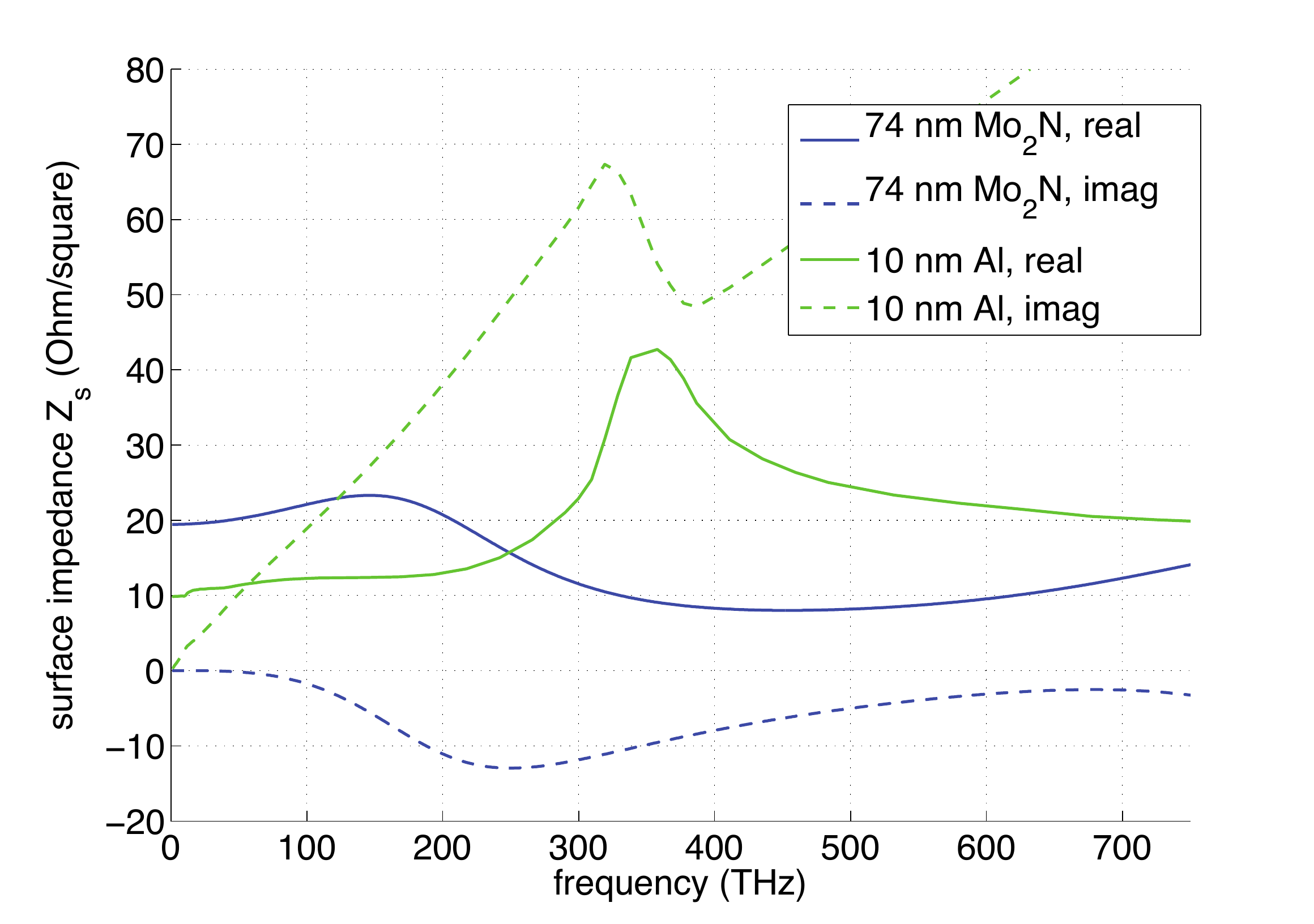}
\caption{\label{fig:MKID_Zs}Surface impedance of two representative MKID thin film materials over the VISIR frequency range 120 - 750 THz ($2.5~\mu\rm m$ - $0.4~\mu\rm m$). The non-constant, reactive impedance presents challenges for efficient optical coupling compared to MKID designs at far infrared frequencies or lower ($< 10$ THz).}
\end{center}
\end{figure}

Optical coupling to VISIR MKIDs efficiently over a wide wavelength range presents a greater challenge than typical for long-wavelength MKIDs.  The large variety of coupling schemes developed for long-wavelength MKIDS fall into two categories:  (1) transmission line coupled, and (2) direct absorption coupled.  In the first category, radiation is collected by an antenna and guided to the MKID by a superconducting transmission line (made of a higher gap superconductor than the MKID).  The MKID is designed to act as a resistive termination (at frequencies above its gap) that matches the characteristic impedance of the optical input line.  For VISIR MKIDs, the optical frequencies are above the gap of any superconductor, so this approach cannot be used.  In the second category, the MKID material is directly illuminated by means of lenses, or placement inside a waveguide.  In the long-wavelength case, the optical frequency is well above the superconducting gap frequency, but far below the inverse of the Drude scattering time.  The thin MKID film then acts as a sheet resistor with a real surface impedance equal to the DC value (typically tens of ohms/square) seen in the normal (non-superconducting) state.  By appropriate choice of the index of refraction of the (transparent) substrate, and use of a back-short, highly efficient optical coupling to the MKID can be achieved over a fractional bandwidth of 30\% or more.  At the much higher frequencies in the VISIR case, MKID materials exhibit a more complex dielectric function.  Fig.~\ref{fig:MKID_Zs} shows the surface impedance at VISIR frequencies for two examples of MKID films, molybdenum nitride and thin aluminum, which we have used at NASA Goddard.  The real part of the impedance is not frequency independent, and the imaginary part is not small.  This is typical of all MKID materials, including TiN, NbTiN, PtSi, WSi.  An optical efficiency near 100\% can be designed in some narrow frequency range by forming an optical cavity involving the MKID layer, its substrate, and auxiliary metal or dielectric films; however, it seems a complex task to achieve $>$ 50\% efficiency simultaneously over 0.4 to 2.5~$\mu\rm m$. Additional complications are (1) the MKID films are not necessarily thin compare to the optical wavelength, (2) one of the favored substrates, single crystal silicon, has its semiconducting gap in the frequency range of interest, and (3) amorphous dielectrics associated with TLS may add noise.  The TiN MKIDs in ARCONS absorb 70\% of the light at $0.4~\mu\rm m$, but only 30\% at $1.0~\mu\rm m$, and microlens arrays are used to focus the light onto the small inductors. \cite{2013PASP..125.1348M} More than one MKID design may be needed in biosignature characterization focal planes to efficiently couple photons from 400~nm - 2.5~$\mu$m. For LUVOIR, this may not be a significant penalty because the coronagraph itself will have limited bandpass, perhaps 10\%, as a consequence of needing to achieve a $ 10^{-11}$ starlight suppression ratio. Nevertheless, decreasing MKID inductor size (to increase spectral resolution) and improving absorption efficiency are important challenges for VISIR MKID development.

\section{Ultra-low Vibration Cooling}\label{sec:cooling}

\subsection{Overview of coolers for \texorpdfstring{$\rm T>4~K$}{T greater than 4 K}}

Stored cryogen systems have been used in the past to provide cooling to observatories and instruments with near zero vibration, but they are impractically massive for missions with lifetimes greater than five years, and have largely been replaced by mechanical cryocoolers.  Cryocoolers are far lighter and have lifetimes limited primarily by their control electronics.   

While there are many types of closed cycle cryocoolers, they generally share several common elements.  All use a working fluid, typically helium, and have a compressor at the high temperature end, followed by a heat exchanger, where the heat of compression is rejected to a radiator.  All have a heat exchanger where the heat from the cold-ward flowing gas is rejected to the warm-ward flowing gas.  In the case of alternating flow (ac) systems, such as Stirling cycle or pulse tube coolers, this is called a regenerator; in the case of continuous flow coolers, such as turbo-Brayton or Joule-Thompson coolers, this is called a recuperator or counterflow heat exchanger.  Finally, in all systems, gas is expanded by various means, then enters a heat exchanger where heat is absorbed at the operating temperature.

\subsection{Linear compressor cryocoolers}
Almost all flight cryocoolers launched to date are based on linear motor driven piston compressors with non-contact clearance seals.  These devices, originally developed in the 1970s and 80s at Oxford University, have virtually unlimited lifetime.  They also have inherently high vibration at their operating frequency, typically 20 to 70 Hz, which unfortunately is in a range that often contains important telescope and instrument structural mode frequencies.  Many flight cryocoolers use a second, co-aligned piston and control electronics to provide active vibration cancellation along the axis of motion, but cancellation is imperfect, partially because the piston force couples into other degrees of freedom.  In most of these coolers, the regenerator and the expansion piston (or pulse tube) are mounted together in a single unit with the compressor, which must be mounted directly to the instrument.  In linear piston driven JT coolers, the alternating flow of a compressor is rectified with a set of reed valves.  This scheme is used on  the JWST/MIRI instrument and the Astro-H/SXS instrument.  The resulting flow, after being cooled by a separate cryocooler, can then be piped many meters to a remote expansion valve.  Although the inherent noise of the flow in the line and in the expansion valve is low, the lines must be directly coupled to the circulating compressor and the cryocooler, and transmit their vibration to sensitive parts of the observatory.  

\subsection{Low vibration cryocoolers}

Because of the known problems with the vibration from linear-piston cryocoolers, alternative coolers with much lower exported vibration force in the critical $0 - 200~\rm Hz$ band have been developed.  Two examples are Joule-Thompson expansion coolers with sorption based compressors (called “sorption coolers” here) and reverse Brayton cycle coolers using miniature turbine compressors and expanders (called “turbo-Brayton coolers” here).  In both cases, the flow is continuous, rather than oscillating, and the compressors can be mounted meters away from the instruments.

\begin{figure}[t]
\begin{center}
\includegraphics[width=0.8\textwidth]{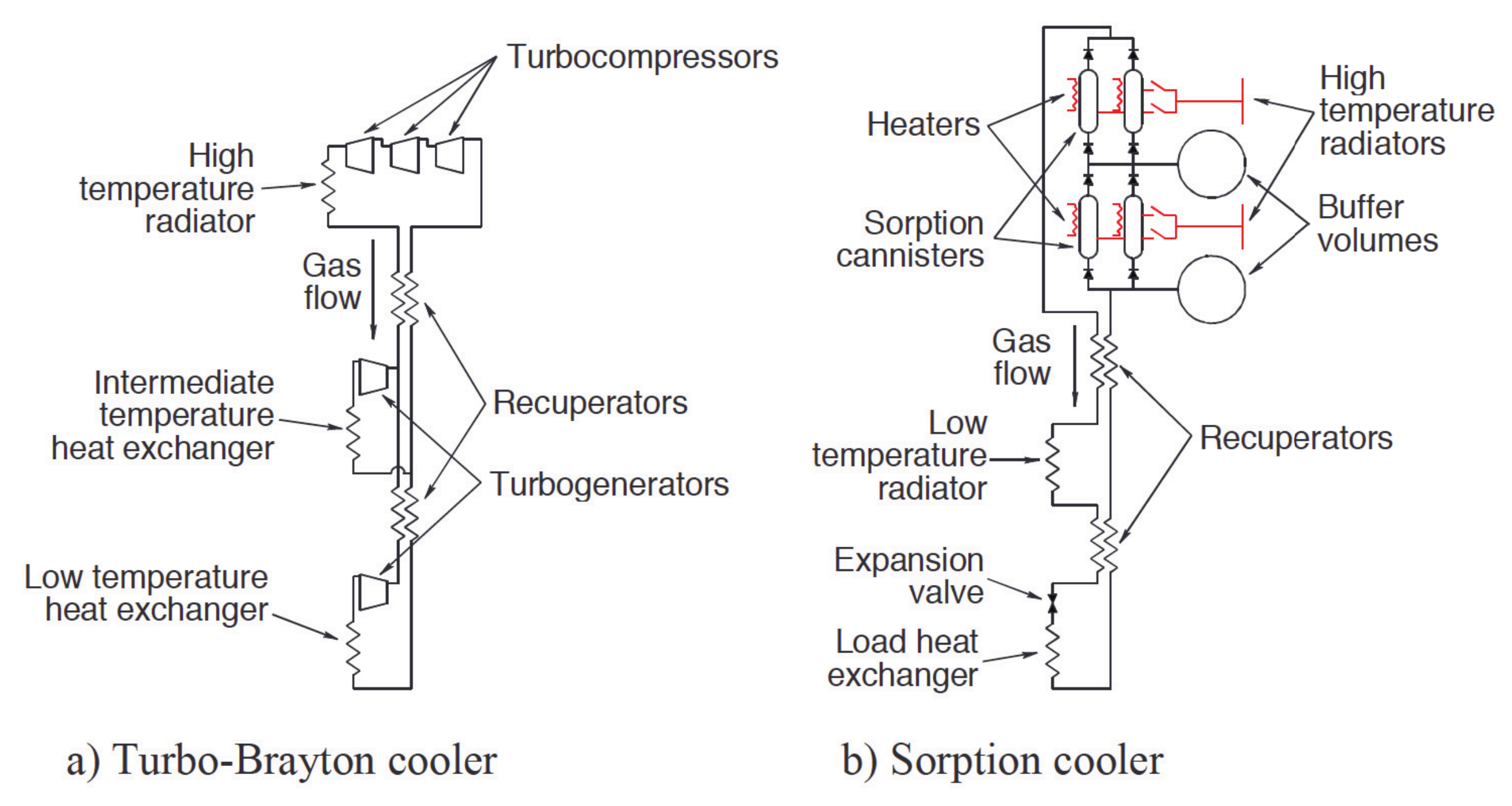}
\caption{\label{fig:cooler_schematics}Simplified schematic diagrams of a) turbo-Brayton and b) sorption coolers. Note that the sorption cooler must reject heat to a low temperature radiator (or a higher temperature cooler) so that the gas is sufficiently cold before reaching the expansion valve, while, in the configuration shown, the turbo-Brayton system has a second intermediate stage which absorbs heat.}
\end{center}
\end{figure}

In the turbo-Brayton cooler (Fig.~\ref{fig:cooler_schematics}a), a motor-driven turbine works on the gas at the warm end, compressing it, and a turbine-driven generator extracts work from the gas as it expands at the cold end.  Because expansion in the turbine ideally approaches an isentropic process, the reverse-Brayton cycle has inherently high efficiency.  The turbines are very small devices, of order several mm, that operate at very high rotational frequency, typically 10 kHz for the compressor and 3 kHz for the expander, which is far above the critical structural mode frequencies for a large telescope such as JWST.  The turbines float on self-actuated gas bearings, and are thus non-contact devices, so the lifetime of the cooler is typically limited only by the rad-hardness of its electronics.  A single stage turbine can produce produce only a relatively modest compression ratio, especially in helium.  This can be offset to some degree by connecting multiple compressors in series, but the compression ratio is typically modest compared to other coolers, and requires a rather large, sophisticated, very high efficiency recuperator.  The turbo-Brayton system can have multiple turbo-expander stages that can absorb heat at multiple temperatures.  While low temperature radiators (in addition to the main warm radiator following the compressor) will improve system efficiency, the cooler can be made to operate without them, and so can have a relatively modest impact on the spacecraft configuration.

Sorption coolers (Fig.~\ref{fig:cooler_schematics}b) are driven by sorption compressors, which are simply beds of material which absorb gas at low temperature, and emit it at high temperature.  To reach the temperatures of interest here, at least a two stages will be needed.  For a hydrogen upper stage, the process is typically chemi-sorption, where the gas reacts with a metal such as LaNiSn to form metal hydrides.  For a helium low temperature stage, this process is typically adsorption onto a highly porous material such as carbon.  While chemi-sorption compressors operate around ambient temperature, carbon sorbents must operate at low temperature ($< 50~\rm K$), and so require a radiator on the cold side of the spacecraft.  At least two beds are required.  At any time, one is cold and absorbing gas at low pressure, while another is warm and emitting gas at high pressure.  Although the process is inherently cyclical, buffer volumes and careful control of cool down and warm-up rates can smooth out pressure fluctuations.  As the sorbent beds switch from cooling to warming and vice-versa, check valves keep the gas moving in one direction.  These are the only moving parts, and they open and close only with the frequency of the heating and cooling of the sorbent beds, which is well below any structural mode frequency for a large telescope, although valve actuation does produce a small impulse with broad frequency content.  Compressors can be staged, and relatively high compression ratios can be achieved, so a relatively simple recuperator can be a used.  At the low temperature end, gas expands isenthalpically through a simple Joule-Thompson valve.  Isenthalpic expansion provides no cooling in an ideal gas, so prior to reaching the expansion valve, it must be cooled well below its region of ideal behavior.  For a system capable of absorbing heat below $\sim 10~\rm K$, a helium cooler would be needed, which will require pre-cooling with a hydrogen stage.  The hydrogen stage would need pre-cooling with a set of staged low temperature radiators, with the coldest radiator at $\sim 50~\rm K$.  Thus, such a cooler would have a significant impact on the spacecraft configuration.

Low vibration coolers have been used in at least two important astrophysics missions.  A turbo-Brayton cooler was installed on the HST/NICMOS instrument during servicing mission 3B to replace a solid nitrogen dewar that had failed.  The cooler was a single stage device that used neon as a working fluid, and provided cooling at 73 K to the NICMOS detectors\cite{Swift:2008}.  Once the cooler reached steady state, it had no detectable effect on HST image quality.  A hydrogen sorption cooler was used on Planck to provide cooling at $<19.5~\rm K$ to a linear compressor-driven helium JT cooler\cite{Planck:2011}.  Its compressor operated between 270~K and 460~K, and a three-stage V-groove radiator was used to provide precooling.  Operation of the sorption cooler caused no detectable noise, although any signal from the sorption cooler would have been minuscule compared to that of the linear compressor.  Since these missions, there have been additional advances in low vibration cooling systems.  Notably, Breedlove, \etal\cite{Breedlove:2014} demonstrated a two-stage turbo-Brayton system that provides 236 mW of cooling at 10 K, and Burger, \etal\cite{Burger:2007} demonstrated a hydrogen/helium sorption cooler that provides 5 mW of cooling at 4.5 K.

\subsection{Sub-Kelvin coolers}
The effectiveness of cooling by the expansion of helium gas drops off rapidly below 1 K, and other physical phenomena must be used to reach deep subKelvin temperatures.  In terrestrial laboratories, where power is effectively free, and gravity provides a natural separation of the $\rm ^3He$-rich and $\rm ^3He$-poor phases of a liquid $\rm ^3He$/$\rm ^4He$ mixture, dilution refrigerators are most commonly used to reach temperatures as low as 0.002 K.  Dilution coolers are based on the entropy of mixing of $\rm ^3He$ in $\rm ^4He$.  An open-cycle dilution refrigerator was developed by the Grenoble group for cooling the HFI detectors on Planck.  The device used $\rm ^3He$ and $\rm ^4He$ stored at room temperature in four large high pressure tanks.  The gas was pre-cooled by the three-stage radiator, the sorption cooler, and the helium JT cooler before reaching the dilution refrigerator, where it provided several hundred nanowatts of cooling at 0.1 K.  The gas lasted 29 months.  The same group is working on a closed cycle dilution refrigerator that relies on surface tension to separate the phases, but so far they have not demonstrated a complete system that will operate without gravity.

\begin{figure}[t]
\begin{center}
\includegraphics[width=0.8\textwidth]{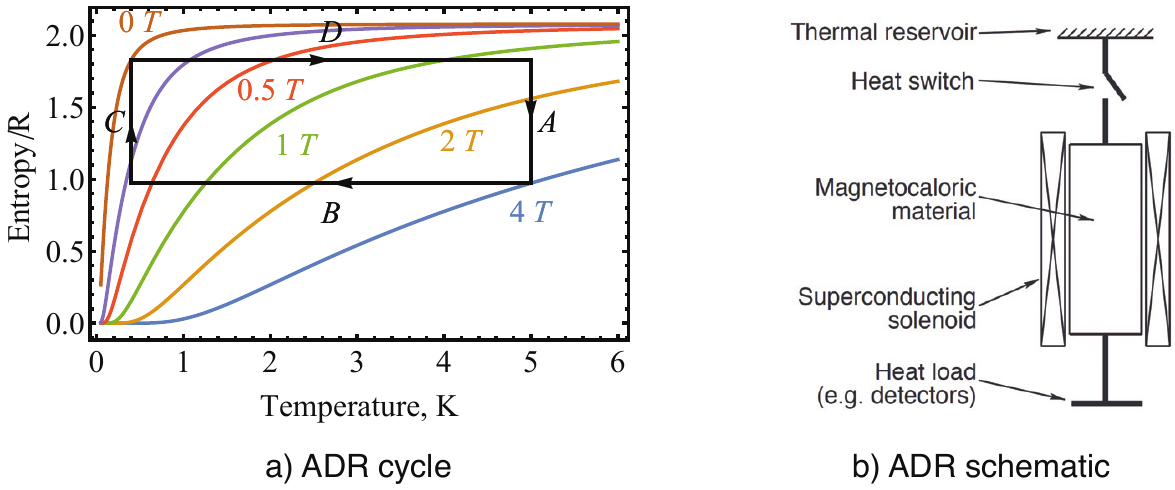}
\caption{\label{fig:ADRcycle}a) Entropy as a function of temperature for a model magnetocaloric material. The superimposed rectangle shows the thermodyanmic path of an ideal adiabatic demagnetization refrigerator. b) Schematic of a basic single-stage ADR. Typically, the device is designed so that recycling (operations D, A, and B in the plot) require less than an hour, while hold time (operation C) is more than a day.  In this example, the thermal reservoir is at 5 K.  The heat switch is only closed during operation A, and the heat of magnetization is transferred to the reservoir in this part of the cycle.}
\end{center}
\end{figure}

The Goddard Space Flight Center has developed magnetic coolers, or Adiabatic Demagnetization Refrigerators (ADRs) for lifting heat from milliKelvin temperatures in a 0-g environment.  Magnetic cooling is based on manipulation of the entropy of paramagnetic compounds with a magnetic field.  Because of their unfilled d and f sub-shells, many rare earth and period 4 transition metal ions have magnetic moments, and have $2J+1$ states, where $J$ is the total angular momentum quantum number.  In the limit of small interaction between ions, these states are degenerate, so the associated entropy is $R \ln(2J+1)$, which at temperatures below $\rm 10 - 15~K$ for most materials is far larger than other entropy terms.  Applying a magnetic field breaks this degeneracy and suppresses the entropy.  At sufficiently low temperature, the interaction between ions also acts to align the moments and suppress entropy.  As the material approaches its ordering temperature, magnetic entropy drops sharply.  Fig.~\ref{fig:ADRcycle}a illustrates this behavior; the solid curves are the entropy as a function of temperature at several values of applied field.  

The rectangle labeled A-B-C-D shows the ideal ADR cycle.  In a single stage ADR, the paramagnetic compound sits in the bore of a superconducting solenoid, thermally connected to a thermal reservoir at the warm end through a heat switch.  In process A, the field is ramped to maximum with the switch closed, so the heat of magnetization is dumped to the reservoir.  In process B, the heat switch is open, so the paramagnetic material is isolated (or adiabatic), and the temperature drops isentropically as the field is reduced until the desired operating temperature is reached.  In C, the field is reduced slowly, at a rate that generates cooling only sufficient to cancel the heat input from the low temperature load, until the field reaches 0.  Finally in D, the field is ramped rapidly up to the reservoir temperature, at which point the heat switch is closed and the cycle repeats.  Note that this is a Carnot cycle, so that in the limit of ideal operation, ADRs have maximum possible thermodynamic efficiency.

NASA Goddard Space Flight Center has built three flight ADR systems.  Two were nearly identical single-stage coolers for the XRS instrument on Astro-E and E2.  They lifted heat from the detector array at 0.060 K to a liquid helium tank at 1.3 K.  The ADRs had a hold time of 33 hours at the detector operating temperature, and had a 1 hour recycle time.  Astro-E2 successfully reached orbit, and the ADR worked flawlessly until the liquid helium ran out.  The third device is a 3-stage ADR for the SXS instrument on Astro-H.  It has multiple operating modes.  In nominal mode, it lifts heat from the detector array at 0.050 K to a liquid helium tank at 1.3 K.  In this mode, the hold time is 49 hours, and the recycle time is only 0.75 hours.  Once the helium runs out, the system provides continuous cooling to the empty tank at 1.5 K, and also cools the detectors to 0.050 K, although with reduced hold time.

As array sizes of low temperature detectors scale up, so does the low temperature heat load.  For standard ADRs, maintaining long hold times requires scaling up the ADR system proportionally.  The Continuous ADR (CADR) circumvents this limitation\cite{Shirron:2004}.  A CADR is a multistage ADR adapted so that the first (coldest) stage stays at the detector operating temperature.  For half of its cycle, this stage operates normally, absorbing heat from the detectors through a controlled ramp-down of its field.  However, as its field approaches zero, the second stage is brought down to a temperature below the operating temperature, and the heat switch is closed.  The first stage must then magnetize to maintain the operating temperature, and in this way transfers the heat it has absorbed to the second stage.  As the field approaches maximum, the heat switch is open, the first stage starts demagnetizing, while the second stage magnetizes up to a higher temperature and transfers its heat to the third stage.  The process can be cascaded, with heat transferred to higher temperature stages, and finally to the heat sink, presumably a cryocooler.  The most obvious benefit of the CADR is that operation is continuous, so there is no interruption of science data taking.  Perhaps more importantly, because operation is continuous, detector operation and ADR operation are decoupled, and the stages can be cycled much more rapidly.  Since the same heat is lifted with each cycle, increasing frequency increases cooling power power per unit mass.  

While a 4-stage CADR lifting heat from 0.035 K to $\sim 5~\rm K$ has been demonstrated, raising the heat sink temperature will enable its use with turbo-Brayton coolers, and greatly ease integration with sorption coolers.  Although magnetocaloric materials will operate effectively above 10~K, compact superconducting magnets made from standard NbTi/copper composite wire cannot reach sufficiently high fields when operating above $\sim 5~\rm K$.  Compact, low current magnets based on $\rm Nb_3Sn$ composite wire can provide sufficient field while operating above 10 K\cite{Tuttle:2008}.  Recently, a simple ADR stage based on such a magnet has demonstrated heat lift from 4 K to 10 K.  With some additional effort, such a stage could be integrated into a CADR that provides heat lift from $\sim 0.035~\rm K$ to greater than 10~K.  It is also possible to design a CADR to lift heat from from temperatures significantly below 35~mK with proper choice of materials.   

\subsection{Suitability of coolers for a LUVOIR mission }
One well developed concept for a LUVOIR mission was ATLAST\cite{Oegerle:2010}.  ATLAST was based largely on extensions of JWST, and because of the similarity, results of detailed structural and optical modeling for JWST provided useful estimates of ATLAST parameters.  To meet its science goals, ATLAST required wavefront stability of 0.01 nm over 10 min.  Feinberg~\etal\cite{Feinberg:2014} considered the sensitivity of wavefront error (WFE) to disturbances.  Sensitivity is worst in the $\rm 20 - 65~Hz$ band containing the tip-tilt modes of the primary mirror segments.  Using results from JWST deployed dynamics modeling, they showed that substantially better isolation from the momentum wheel assembly disturbances would be required, and argued that this could be achieved using a non-contact linkage between the spacecraft bus, including the sunshield, and the telescope.  For both the turbo-Brayton and sorption coolers, the compressors could be mounted on the spacecraft side.  In both cases, exported disturbances would be far less than those of the momentum wheel assemblies.  However, flow lines, heat exchangers, and expansion valves (for the sorption cooler) or expansion turbines (for the turbo-Brayton cooler) would need to be mounted on the telescope structure.  

The MIRI JT cooler has similar flow lines, heat exchangers, and an expansion valve. Using the same deployed dynamics model, the JWST team examined the sensitivity of WFE to disturbance caused by turbulent flow in the MIRI cooler.  Using computational fluid dynamics, they derived the power spectral density (PSD) of force inputs at the various mounting points to the structure.  These are bounded by $\rm 7~\mu\rm N/\sqrt{Hz}$.  The resulting WFE, integrated up to 200 Hz, is $\rm \sim3~nm$.   Thus, to be a small part of the ATLAST WFE budget, these disturbances would need to be reduced by at least 3 orders of magnitude.  Similar computational fluid dynamics calculations were recently done to determine the noise generated by various components of a turbo-Brayton cooler.  While the worst noise generators, such as step changes in cross-sectional area, would be avoided in an ultra-low vibration cooler, even a relatively minor obstacle, an over-penetrated weld in a straight pipe, produced a force PSD of $\rm \sim0.3~\mu\rm N/\sqrt{Hz}$.  

One potential way to achieve extremely low vibration levels during exoplanet observations would be to use a thermal storage device to provide cooling, and simply switch off the cooler during this period. For example, a reservoir of evaporating liquid helium could absorb heat from the instrument during observations, and the gas could be collected in a tank.  Between observations, the cooler could be turned on to re-liquefy the gas.  However, heat loads on the instruments could be high. For example, in the ATLAST concept, the telescope structure which surrounded and supported the instrument was controlled at ambient temperature.  This, combined with the expected long observation period (up to days), means that such a thermal storage unit would need to be large and heavy.  Since it operates with a limited duty cycle, the cooler would also need to be larger and heavier, and its required power correspondingly larger.  Furthermore, at the 10 pm WFE level, it may be difficult to mitigate the impact on dimensional stability of switching between two modes, one in which the cooler lines are drifting up in temperature, and one in which they are cold.  A better approach, and one in line with the overall architecture of ATLAST, would be to develop the technology to allow the cooling system to be maintained in a steady state.

Advancing turbo-Brayton and sorption coolers so that the exported vibrations are in the single-digit $\rm nN/\sqrt{Hz}$ will require a significant technology investment.  Careful design and fabrication of the entire flow path to eliminate any sharp changes in curvature and including the line stiffness in the computation of forces will likely lead to more than an order of magnitude reduction. However, to reach the desired levels, it may be necessary to ensure laminar flow with no regions of flow separation throughout the flow path.  Laminar flow without separation is completely steady state, so in principle should produce no vibration.  However, flow lines will need to be substantially larger to keep the Reynolds number below the critical value.  Other modifications may also be necessary.  At the expansion valve outlet in hydrogen sorption coolers, the fluid is typically in two-phase flow, which is generally noisy.  It may be necessary to avoid this, although it will impact performance.  For turbo-Brayton coolers, imbalance in the turbo expander rotor causes a disturbance at the rotational frequency of the turbine.  With current rotor balancing technology, the disturbance amplitude is typically hundreds of mN.  Although the impact of a force input at these high frequencies is less well understood, clearly a very large isolation factor is required.  One possibility may be to follow the approach of Aldcroft~\etal\cite{Aldcroft:1992}, who designed and built a 6 stage, 6 degree-of-freedom isolator with at least 250 dB of attenuation in the desired frequency band. 

ADRs have no moving parts and are generally considered to be zero-vibration devices.  However, the stresses in the magnets cycle up to fairly high levels as the fields ramp up and down, and it will be necessary to determines if this generates any disturbances at the relevant level.  This points out an important technology need: experimental techniques for detecting extremely low disturbance forces.  Such techniques will be necessary for other telescope components.

\section{Summary}\label{sec:summary}

We have discussed a broad suite of detector and cooling technologies for biosignature characterization using future space observatories such as LUVOIR and the Habitable-Exoplanet Imaging Mission. For easy reference, Tab.~\ref{tab:summary} summarizes some of these technologies, and the challenges with reference to the state-of-the-art. 

\begin{table}[t]
\caption{Summary of where further work is desirable} 
\label{tab:summary}
\begin{center}
\includegraphics[width=.8\textwidth]{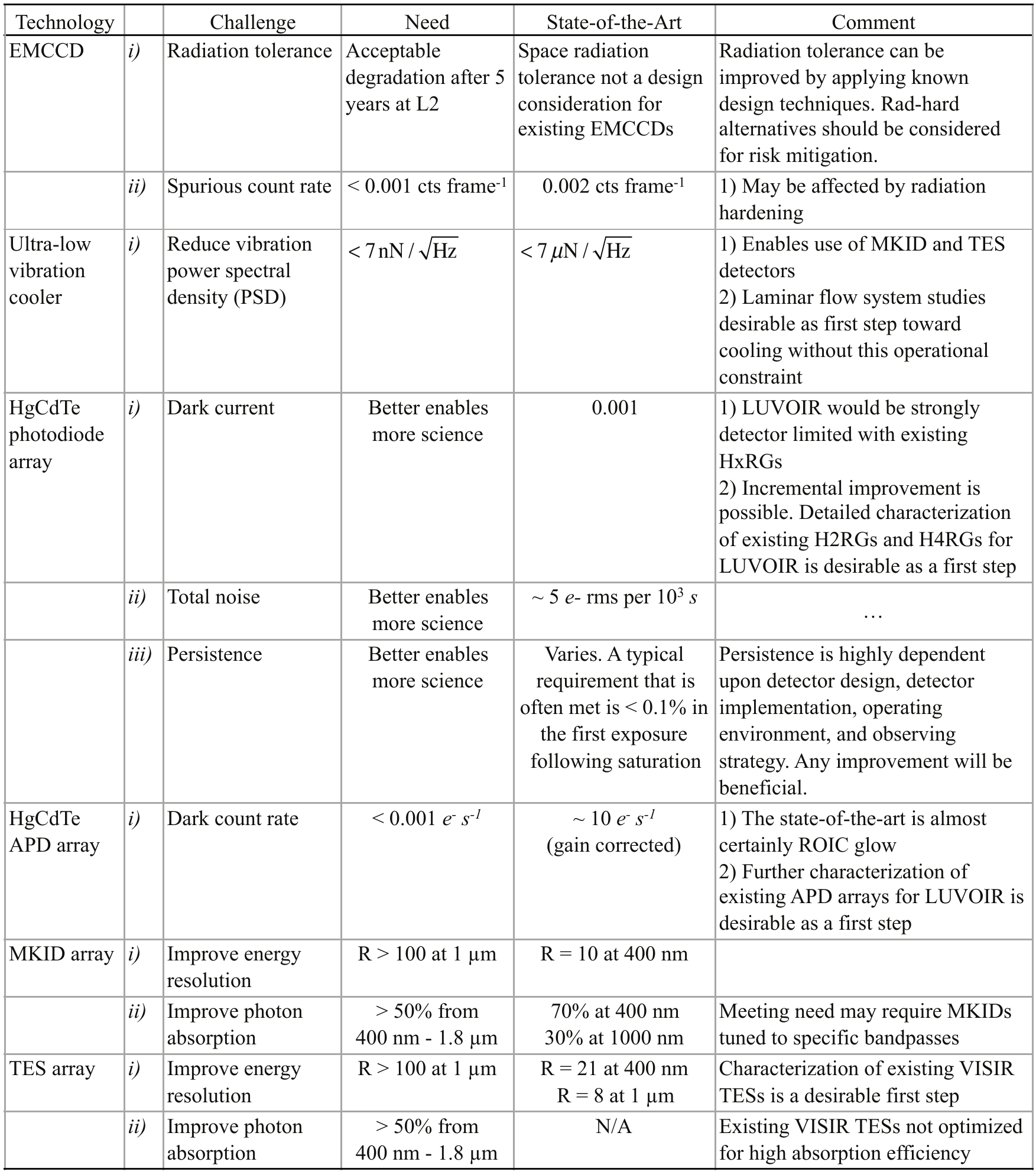}
\end{center}
\end{table}

For EMCCDs, improving radiation tolerance is arguably the greatest need. As is discussed in Sec.~\ref{sec:emccd}, radiation tolerance was not a design consideration for current generation EMCCDs. One should not be surprised to see the radiation induced performance degradation that is typical for n-channel CCDs in space (\eg charge transfer efficiency degradation), and other artifacts that may be revealed at sub-electron noise levels (CIC is one example, but surprises are also possible). For LUVOIR and/or a Habitable Exoplanet Imaging Mission, we believe it would be wise to apply known CCD radiation hardening design features and fabrication processes to EMCCDs.\cite{Burt:2009ct} For risk mitigation, it may also make sense to explore similar detector architectures that promise greater radiation tolerance.

It would also be desirable to improve CIC in EMCCDs, for which the current state-of-the-art is already close to ``good enough'' when new. For CIC, we believe that incremental improvements in operation and design hold good promise for meeting the need on the relevant timescale.

There is still some room from improvement in near-IR photodiode arrays similar to the HxRGs that are being used for JWST, Euclid, and WFIRST. Although the current architecture seems unlikely to function as a single photon detector, significant incremental improvement (perhaps factors of 2-3 reduction in read noise) may be possible. A reasonable first step would be detailed characterization of existing HxRGs aimed at separating out the different contributors to the noise (photodiode, resistive interconnect, pixel source follower, other amplifiers, \etc). Near-IR APD arrays like those made by Selex may also be promising if the ``dark current'' can be reduced to $<0.001~e^-~s^{-1}$. The $\sim 10~e^-~s^{-1}$ gain corrected ``dark current'' of current devices is almost certainly dominated by ROIC glow, but there may still be significant work required to go from the to-be-determined leakage current of these devices to the $<0.001~e^-~s^{-1}$ that is needed.

Superconducting MKID and TES arrays already function as single photon detectors and both have already been used for VISIR  astronomy. Use of these technologies by LUVOIR is contingent upon developing ultra-low vibration cooling. If ultra-low vibration cooling is available, then the challenges for both MKID arrays and TES microcalorimeter arrays are similar. Higher energy resolution and better photon coupling efficiency are needed. If ultra-low vibration cooling is not available, then we believe MKID and TES microcalorimeter arrays may still be attractive for a starshade-based Habitable Exoplanet Imaging Mission because they would offer nearly quantum limited performance.

With specific regard to MKID arrays, further work should include the development of VISIR MKID arrays with designs targeting the energy resolution and optical efficiency required for biosignature characterization. Several areas of investment will be needed.  One expects significant resolution improvements over the state-of-the-art in the near-term from the development of broadband parametric amplifiers with nearly quantum-limited sensitivity, and from switching to MKID materials with greater uniformity in thin-film properties that will eliminate position-dependent broadening of the measured photon energy. In addition, reaching Fano-limited energy resolution will likely require designs that reduce VISIR MKID inductor volume by a factor on the order of 30 from current devices designed for the optical background in ground-based instruments, while at the same time managing to improve optical efficiency.  Achieving high enough optical efficiency over the broad LUVOIR bandwidth will be challenging given the non-constant, reactive complex resistivity of MKID materials at VISIR frequencies.  However, even achieving the Fano-limit with currently favored MKID materials (transition temperature $\approx$ 1 K) will not be sufficient to reach biosignature characterization goals.  Either VISIR MKIDs (and cooling systems) will need to be developed with lower $T_{c} \approx$  0.17 K (operating T $\approx$ 20 mK) in order to give a better Fano-limit, or else effort will be needed to optimize the TKID (membrane) style of detector in order to circumvent the Fano limit for VISIR MKIDs.

Both MKIDs and TESs require ultra-low vibration cooling for use in a LUVOIR. For a Habitable Exoplanet Imaging Mission, the vibration requirements may be less stringent. For these technologies to be viable in all biosignature characterization mission architectures, We recommend the development of prototype technology for ultra-low vibration coolers. As a first step, studies are needed to examine the feasibility of a laminar-flow system, including a detailed computational effort to determine whether flow separation can be avoided.  Once feasibility has been established, the most immediate need will be for techniques that can be used to verify the computational models in prototype components at the required nN levels.

\appendix

\section{Count Rate of Energy Resolving vs Conventional Pixels}\label{sec:count-rate}

In Sec.~\ref{sec:starved}, we assert that if an energy resolving detector were to be used for non-dispersive imaging spectrometry, then the count rate per pixel would be about $100\times$ the count rate per pixel of a conventional IFU spectrograph. The order of magnitude derivation is as follows.

Stark~\etal~(2015)\cite{Stark:2015er} studied space observatory exoEarth yields to suggest lower limits on telescope aperture size. Their study required them to model the performance of both a conventional IFU spectrograph and a non-dispersive imaging spectrograph. Tab.~\ref{tab:stark} lists their key assumptions.

\begin{table}[t]
\caption{Model Assumptions} 
\label{tab:stark}
\begin{center}
\includegraphics[width=0.66\textwidth]{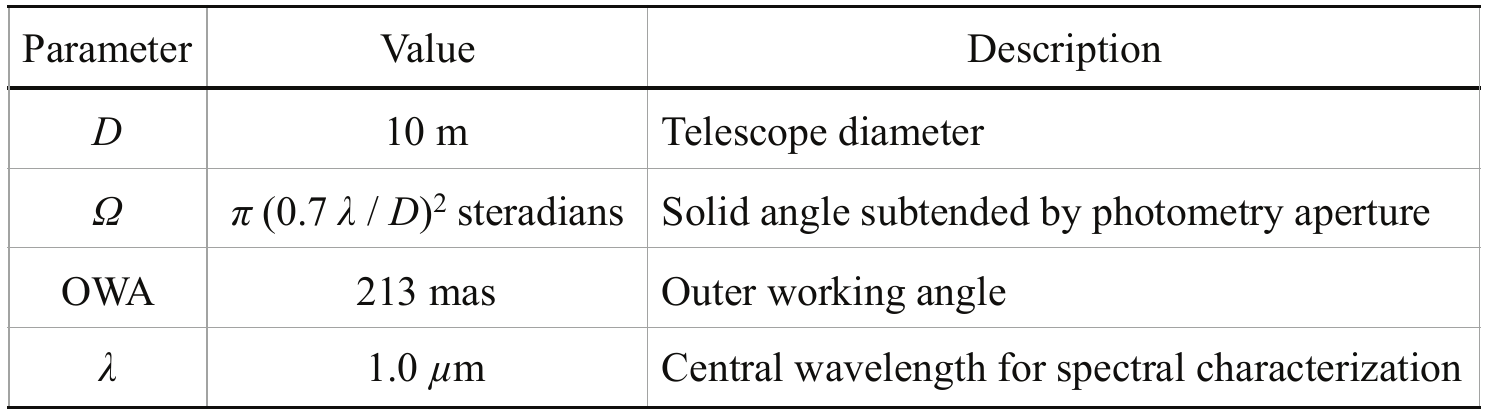}
\end{center}
\end{table}

Following Stark, the photometer aperture, $\Omega$, maps onto 4 energy resolving pixels in the non-dispersive imaging spectrometer. The required number of energy resolving pixels is therefore,

\begin{equation}\label{eq:energy-resolving}
n_{\rm pix} = 4\frac{\left(2~\rm OWA\right)^2}{\Omega} = 1108~{\rm pixels}.
\end{equation}

In the IFU implementation, the photometer aperture maps onto 4 lenslets. Stark furthermore maps each lenslet onto 6 conventional pixels, 3 in the spatial dimension by 2 in the spectral dimension, yielding 24 pixels per spectral resolution element. He assumed a 20\% bandpass and $R=50$ per spectral ``channel'', yielding 240 conventional pixels per photometric aperture. In this article, we have adopted $R=100$ as being better matched to characterizing \oxygen. Following Stark, but requiring $R=100$, yields 480 conventional pixels per photometric aperture. With these assumptions, Eq.~\ref{eq:energy-resolving} becomes

\begin{equation}\label{eq:conventional}
n_{\rm pix} = 480\frac{2~\rm OWA^2}{\Omega} = 133,004~{\rm pixels}.
\end{equation}

If we assume that the overall throughput is about the same in the two implementations, then the same light is being spread over $120\times$ more pixels in the conventional IFU spectrograph than in the non-dispersive imaging spectrometer. To within the uncertainties, this implies that the count rate per pixel will be about $100\times$ higher in the energy-resolving detector than in the conventional detector.

\acknowledgments 
We wish to thank Brendan Crill and Warren Holmes of NASA Jet Propulsion Laboratory and Matthew Greenhouse of NASA Goddard Space Flight Center for carefully reading the entire manuscript and providing invaluable comments. We wish to thank the referee for several helpful comments and suggestions that have improved the manuscript. This work was supported by a NASA Goddard Space Flight Center Internal Research and Development (IRAD) award to develop space coronagraph detector technology and a NASA Goddard Space Flight Center Science and Exploration Directorate Science Task Group award entitled, "Life Finder Detectors".

\bibliography{article}   
\bibliographystyle{spiejour}   


\vspace{2ex}\noindent\textbf{Bernard J. Rauscher} is an experimental astrophysicist at NASA Goddard Space Flight Center. His research interests include astronomy instrumentation, space detector systems, extragalactic astronomy and cosmology, and most recently the search for life on other worlds. Rauscher's work developing detector systems for the James Webb Space Telescope has been recognized by a shared Congressional Space Act award and NASA's Exceptional Achievement Medal.

\vspace{2ex}\noindent\textbf{Edgar R. Canavan} is an aerospace engineer at NASA Goddard Space Flight Center.  His research interests include magnetic cooling and the properties of materials and devices at low temperatures.

\vspace{2ex}\noindent\textbf{Harvey Moseley} is a Senior Astrophysicist at NASA Goddard Space Flight Center. Moseley has received many awards for his work, including SPIE's 2013 George W. Goddard Award. The citation reads in part, \emph{in recognition of his extraordinary inventions of superconducting imaging arrays for astronomy, ranging from sub-millimeter bolometers to energy sensitive X-ray microcalorimeters, and even dark matter detectors [...]}.

\vspace{2ex}\noindent\textbf{Thomas R. Stevenson} is an electronics engineer at NASA Goddard Space Flight Center.  His research interests include superconducting properties of materials and devices, and development of photon detectors for astrophysics applications in spectral regions ranging from microwave, sub-mm, and far infrared, to x-rays.

\vspace{2ex}\noindent\textbf{John E. Sadleir} (Jack) is a condensed matter physicist in the Detector Systems Branch at NASA Goddard Space Flight Center.  His research focuses on cryogenic detectors for particle physics, cosmology, and astrophysics applications.

\vspace{1ex}

\listoffigures
\listoftables

\end{spacing}
\end{document}